\documentclass[journal,12pt,onecolumn,draftclsnofoot,]{IEEEtran}

\usepackage{textcomp}
\usepackage{xcolor}
\usepackage{graphicx}
\usepackage{amsmath,amssymb,amsfonts}
\usepackage{algorithm}
\usepackage{algorithmic}
\usepackage{cite}
\usepackage{lipsum}
\usepackage{epstopdf}
\usepackage{stfloats}
\usepackage{bm}
\usepackage{xcolor}
\usepackage{colortbl,booktabs}

\newenvironment{proof}{{\indent \it Proof:\;\,}}{\hfill $\blacksquare$\par}

\begin{document}
\title{QoS-Oriented Sensing-Communication-Control Co-Design for UAV-Enabled Positioning}

\author{Zijie~Wang,
        Rongke~Liu,~\IEEEmembership{Senior Member,~IEEE,}
        Qirui~Liu,
        and~Lincong~Han
\thanks{This work was supported by the Beijing Municipal Science and Technology Project (Z181100003218008).}
\thanks{Z. Wang, R. Liu, Q. Liu and L. Han are with the School
of Electronic and Information Engineering, Beihang University, Beijing 100191,
China (e-mail: wangmajie@buaa.edu.cn; rongke\_liu@buaa.edu.cn).}
}

\markboth{Journal of \LaTeX\ Class Files,~Vol.~14, No.~8, August~2015}%
{Shell \MakeLowercase{\textit{et al.}}: Bare Demo of IEEEtran.cls for IEEE Journals}

\maketitle

\begin{abstract}
Unmanned aerial vehicle (UAV)-enabled positioning that uses UAVs as aerial anchor nodes is a promising solution for providing positioning services in harsh environments. In previous research, the state sensing and control of UAVs were either ignored or simply set to be performed continuously, resulting in system instability or waste of wireless resources. Therefore, in this article, we propose a quality-of-service (QoS)-oriented UAV-enabled positioning system based on the concept of sensing-communication-control (SCC) co-design. We first establish the mathematical models of UAV state sensing and control. Then, the influence of sensing scheduling and transmission failure on UAV stability, as well as the performance of positioning services in the presence of UAV control error, are analyzed. Based on these models and analysis results, we further study the problem of minimizing the amount of data transmitted by optimizing the sensing scheduling and blocklength allocation under the condition of satisfying each user's demand. Finally, an efficient scheme is developed to solve this mixed-integer nonlinear problem. Numerical results show that the proposed system could work efficiently and meet users' requirements. In addition, compared with two benchmark schemes, our scheme reduces the failure rate or resource consumption of positioning services by more than 76.2$\%$ or 82.7$\%$.
\end{abstract}


\IEEEpeerreviewmaketitle

\section{Introduction}
\subsection{Motivation}
\IEEEPARstart{A}{s} the efficiency and intelligence of industrial manufacturing and people's daily lives has been improved continuously by location-based services (LBS) like logistic tracking and mobile marketing, positioning technology is playing an increasingly important role in this mobile information era \cite{LBS_Intr}. Therefore, ubiquitous positioning has been recognized as an essential service and an enabling technology by both the current fifth generation (5G) wireless networks and the future six generation (6G) communications \cite{Pos_5G,Pos_6G}. Unfortunately, conventional wireless positioning technologies represented by global navigation satellite system (GNSS) and cellular-based positioning may suffer severe performance degradation in some harsh environments like dense urban and fail to meet application requirements. For the widely used GNSS systems, their signals are easily blocked or reflected by buildings or rugged terrain, resulting in multipath or none-line-of-sight (NLoS) propagation that cause large position errors \cite{GNSS_NLoS}. Moreover, due to frequent signal blockages, the geometry of satellites available to users is often unsuitable for positioning \cite{GNSS_Geo}. Similarly, the above problems, including NLoS propagation, insufficient number and unsatisfactory geometry of base stations (BS) can also be found in cellular-based positioning, which severely affect the positioning accuracy \cite{Cell_Num,Cell_NLoS,Cell_Geo}. Thus, we need a better platform to provide wireless positioning services in harsh environments.

Due to their high operational flexibility and fully controllable mobility, unmanned aerial vehicles (UAV) have recently received considerable attention from the research community and may help create a whole new paradigm for wireless networks \cite{UAV_Para}. In the past few years, researchers have studied the potential of UAVs as aerial BSs, relays and data collectors to provide services for ground users or assist terrestrial networks \cite{UAV_ABS,UAV_Relay,UAV_Data}. It is noteworthy that in addition to their use for communication, UAV could also be a promising platform for positioning \cite{UAV_Pos}. Unlike satellites operating in fixed orbits or BSs that cannot be moved, UAVs with high maneuverability can fly close to users, making the number of available anchor nodes sufficient for positioning \cite{UAV_Avai}. Moreover, UAVs are more likely to establish line-of-sight (LoS) links with ground users than terrestrial BSs due to their relatively high flying altitudes \cite{UAV_Para}. In addition, by considering UAV-user geometry in the deployment of UAVs, the problem of poor geometry common in conventional positioning technologies can also be largely avoided \cite{UAV_Geo}. Therefore, UAV-enabled positioning that uses UAVs as aerial anchor nodes could be a good choice for positioning in challenging environments.

As typical automatic systems, the operation of UAVs or UAV swarms requires appropriate design and coordination of sensing, communication and control functionalities \cite{SCC_Sur}. In existing systems, these three functionalities were commonly designed and implemented separately, which was previously thought to be conducive to ensuring compatibility among different functionalities \cite{SCC_Pre}. Recently, many studies pointed out that considering the co-existence and cooperation of multiple functionalities at the beginning of system design can help improve system performance or resource efficiency \cite{SCC_Pre,SCC_Ben_1}. Thus, sensing-communication co-design (also called integrated sensing and communication) that enables the network and users to sense their dynamic states or the surrounding environment has been widely viewed as a key driver in future 6G networks \cite{SCC_SeCo}. The capability of state sensing helps to reduce the energy or resource consumption in wireless networks, and location-aided resource scheduling is a good application example \cite{SCC_Ben_2}. Similarly, control and communication functionalities can also be considered jointly to provide extra design freedom for future automatic systems. Recent research indicated that control-communication co-design could address problems existing in a single functionality, or even create new applications by taking advantage of the imperfections of each functionality \cite{SCC_Pre}. Moreover, if the communication functionality is designed based on system's actual needs for control, the current stringent requirements on communication latency and reliability can be greatly relaxed, thereby reducing unnecessary resource consumption \cite{SCC_XUR}. Therefore, it is very promising to develop sensing-communication-control (SCC) co-design solutions for UAV-related applications \cite{SCC_Ben_2,SCC_UAV}.

\subsection{Related Work}
As society's demand for ubiquitous positioning services increases continuously, UAV-enabled positioning has recently become a hot topic for research due to its high flexibility and adaptability in harsh environments. Sallouha \emph{et al.} \cite{UAV_RSS_1} proposed a UAV-enabled positioning system that utilizes the received-signal-strength (RSS) technique to locate ground users in urban areas and derived the corresponding theoretical performance limits. In \cite{UAV_RSS_2}, the UAV trajectory in the same system was optimized to find a good balance between the accuracy of RSS-based two-dimensional (2-D) positioning and UAVs' propulsion energy consumption. Wang \emph{et al.} \cite{UAV_Avai} introduced the time-difference-of-arrival (TDoA) approach into UAV-enabled positioning and realized high-accuracy 3-D localization by exploiting the vertical diversity of UAV platforms. Moreover, in \cite{UAV_Geo}, the reliability of UAV-enabled positioning services in mountainous areas was modeled and enhanced. In these studies, UAVs were treated as anchor nodes with perfect knowledge of their own positions. In practice, UAVs' positions are commonly obtained through state sensing, and the inevitable sensing errors will cause uncertainty in anchor position information \cite{UAV_Unc_1,UAV_Unc_2}. Thus, the authors in \cite{UAV_Unc_1} studied the problem of UAV self-localization and evaluated the impact of UAV position uncertainty on positioning performance. Furthermore, Liu \emph{et al.} \cite{UAV_Unc_2} proposed a deployment optimization method considering UAV position uncertainty to improve the accuracy of UAV-enabled positioning. These two studies assume that UAVs can hover stably at fixed positions, which is not true in practical applications due to unintentional movements caused by environmental factors like wind. Hence, the authors in \cite{UAV_Unc_3} quantitatively analyzed the influence of the instability and mobility of UAV platforms on UAV relative localization. However, in this study, the anchor position uncertainty in different directions caused by UAV instability was simply modeled as independent Gaussian random variables with the same variance, which may be too optimistic for practice. Therefore, in order to ensure the practicability of UAV-enabled positioning, both the state sensing error and control error should be taken into consideration in system design, and they should be modeled objectively according to the system dynamics model and control policy.

In terms of the co-design of different functionalities, several successful attempts have been made in recent years. In \cite{SCC_RAC_1}, the authors achieved the co-existence between radar and communication systems through cooperative spectrum sharing, and studied the problem of cross interference management. The two systems studied in this research still work separately on different platforms for different users. No other interactions between the sensing and communication functionalities were considered in this study except for the cross interference. Liu \emph{et al.} \cite{SCC_RAC_2} integrated radar sensing and millimeter wave (mmWave) communication functionalities on a multiple-input multiple-output (MIMO) system, and designed the corresponding frame structure. Nevertheless, in this study, the users or targets of the two functionalities are different, which means that a single user cannot benefit from both functionalities. In \cite{SCC_VCR}, the vehicle motion predicted through radar sensing was used to assist the reception of downlink communication signals. Unfortunately, controls issues are not considered in this study, making the vehicle dynamics model used less practical. Mei \emph{et al.} \cite{SCC_ACC_1} and Gonz\'alez \emph{et al.} \cite{SCC_ACC_2} introduced the communication-control co-design into vehicle-to-vehicle (V2V) networks, and developed resource scheduling algorithms to ensure the string stability of vehicle platoons. In these two studies, the sensing of each vehicle's kinematic status was assumed to be performed perfectly without failures or errors, which is not true in practical applications. In \cite{SCC_CAC_1} and \cite{SCC_CAC_2}, control-aware communication was used to improve the resource efficiency of wireless control systems. These two studies focused on the stability and resource consumption of a control system, rather than the system's ability to perform the required tasks. In fact, a system may not need to be very stable to provide services that satisfy users' demands, and stringent requirements on stability may lead to a waste of resources. From the above analysis, it can be concluded that if we want to improve the performance or efficiency of an automatic system through the joint design of multiple functionalities, the following requirements need to be met in system design.
\begin{itemize}
\item Essential functionalities including sate sensing, communication and control should be comprehensively considered.
\item The imperfections of each functionality should be modeled objectively.
\item The main consideration should be the performance of the services provided for users.
\end{itemize}

\subsection{Main Contributions}
In this article, a quality-of-service (QoS)-oriented UAV-enabled positioning system is proposed based on the idea of SCC co-design to improve the resource efficiency of positioning services while ensuring that users' requirements on positioning accuracy are met. To evaluate objectively the performance of our system in practical applications, we consider the impact of UAV position uncertainty caused by the state sensing error and control error on positioning. Specifically, the mathematical models of UAV state sensing and control in two operation modes, ``open-loop'' and ``closed-loop'', are first established. Based on these models, we further derive the covariance matrices of UAV control errors in different operation modes, as well as the mean-square error (MSE) of positioning services in the presence of UAV position uncertainty. It is found that the quality of positioning services is mainly determined by UAVs' operation modes, while the latter depends on the scheduling of UAV state sensing and successful transmissions of sensing data with finite blocklength, resulting in the coupling among sensing, communication and control functionalities. Then, we try to reduce the resource consumption for positioning services by reducing the frequency of state sensing and the blocklength for data transmission, while ensuring that users' QoS demands are satisfied. This work is formulated as a mixed-integer nonlinear optimization problem, and an efficient scheme is developed to solve it.

In a nutshell, the major contributions of this article are listed as follows:
\begin{itemize}
\item Different from previous studies that treated UAVs as perfect anchor nodes, we consider UAV position uncertainty caused by the state sensing error and control error in system design and analysis. In particular, the imperfections of each functionality are modeled objectively, making our system more practical than existing ones.
\item We derive the covariance matrices of UAV position uncertainty and the MSE of positioning services in different operation modes. These derivations can be used as a general framework for the design and analysis of UAV-enabled positioning systems, and provide guidance for future research on SCC co-design like resource allocation.
\item When solving the problem of sensing scheduling and blocklength allocation, we develop a scheme that converts the constraint on the positioning MSE of each user into the constraint on each UAV's control error. This scheme decouples the control of each UAV, and is therefore suitable for large-scale UAV-enabled positioning systems.
\end{itemize}

Numerical results demonstrate that our proposed system and scheme achieve the desired results in terms of QoS and resource efficiency. Moreover, compared with two benchmark schemes, the proposed scheme achieves significantly lower failure rate or better resource efficiency. To the best of our knowledge, this work is the first to introduce the concept of sensing-communication-control co-design into UAV-enabled positioning.

\emph{Notations:} In this article, scalars are denoted by italic letters ($x$). Lowercase and uppercase boldface letters (${\bf{x}}$ and ${\bf{X}}$) are used to denote column vectors and matrices, respectively. ${\mathbb{R}^{N \times 1}}$ represents the space of $N$-dimensional real-valued vectors. The superscript $T$ indicates the transpose operation (${{\bf{X}}^T}$) and superscript $-1$ indicates matrix inverse (${{\bf{X}}^{ - 1}}$). ${\left\|  \cdot  \right\|_2}$ and ${\mathop{\rm tr}\nolimits} \left(  \cdot  \right)$ denote the Euclidean norm and matrix trace, respectively. $E\left\{  \cdot  \right\}$ represents the statistical expectation operator and $P\left(  \cdot  \right)$ indicates the probability. ${\mathop{\rm diag}\nolimits} \left(  \cdot  \right)$ and ${\mathop{\rm blkdiag}\nolimits} \left(  \cdot  \right)$ denote the diagonal and block diagonal matrices, respectively. ${{\bf{I}}_N}$ is the $N \times N$ identity matrix and ${{\bf{0}}_{N \times M}}$ represents the $N \times M$ all-zero matrix. ${\cal N}\left( {{{\bf{0}}_{N \times 1}},{\bf{Q}}} \right)$ denotes the $N$-dimensional Gaussian distribution with zero mean and covariance matrix ${\bf{Q}}$. $\dot v$ indicates the first-order derivative of the time-dependent function $v$ with respect to time.

\section{System Design}
\begin{figure}[!t]
\centering
\includegraphics[height=2.9in,width=3.45in]{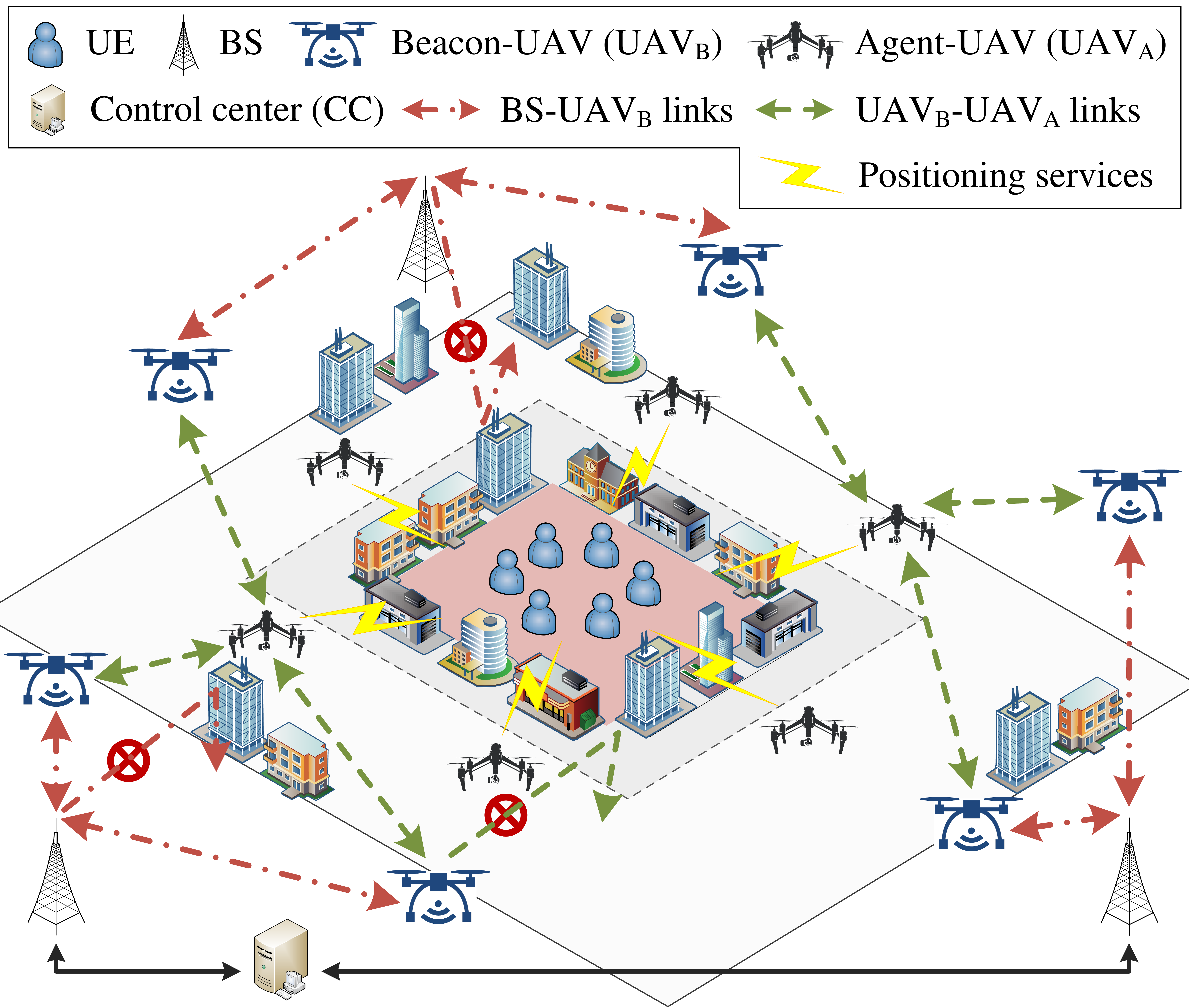}
\caption{Proposed UAV-enabled positioning system.}
\label{fig_1}
\end{figure}

In this article, as shown in Fig. 1, we consider a scenario where multiple ground users requiring positioning services are located in a challenging environment. In such an environment, conventional technologies such as GNSS and terrestrial cellular-based positioning fail to meet users' requirements due to their drawbacks discussed in the previous section. Therefore, multiple low-altitude UAV platforms are employed as aerial anchor nodes to undertake the task of locating ground users. These UAVs can be classified into two groups according to their functions in the proposed system, namely the ``agent UAV (UAV$_{A}$)'' and ``beacon UAV (UAV$_{B}$)''. Among them, UAV$_{A}$ are responsible for providing positioning services to users. Since UAV$_{A}$ are deployed close to ground users, the air-to-ground (A2G) links between UAV$_{A}$ and user equipment (UE) are dominated by LoS components. However, due to their long distance from terrestrial BSs, UAV$_{A}$ are unable to determine their own locations through cellular-based positioning. Thus, UAV$_{B}$ whose locations have been accurately estimated and remain unchanged are used to provide UAV$_{A}$ with relative range measurements required for state sensing.

The proposed system shown in Fig. 1 consists of ${N_A}$ UAV$_{A}$, ${N_B}$ UAV$_{B}$ and $M$ UE. UAV$_{A}$ and UAV$_{B}$ are denoted by sets ${{\cal N}_A} = \left\{ {1, \cdots ,{N_A}} \right\}$ and ${{\cal N}_B} = \left\{ {1, \cdots ,{N_B}} \right\}$, respectively. The $j$-th UAV$_{A}$ ($j \in {{\cal N}_A}$) is deployed at a carefully selected ``hovering point (HP)'', which can be denoted by the horizontal coordinates ${\bf{q}}_j^ \circ  \in {\mathbb{R}^{2 \times 1}}$ and height ${h_v}$. The 3-D coordinates of $j$-th UAV$_{A}$'s HP are denoted by ${\bf{q}}_j^{ \circ ,3D} \!=\! {\left[ {{{\left( {{\bf{q}}_j^ \circ } \right)}^T},{h_v}} \right]^T} \!\in {\mathbb{R}^{3 \times 1}}$. As mentioned in the previous section, in practice, UAV$_{A}$ are unable to hover stably at fixed positions, and can only maintain themselves in close vicinity of the selected HPs through sensing and control functionalities. Moreover, it is assumed that UAV$_{A}$ can maintain the preset altitude (${h_v}$) utilizing their onboard sensors like barometric altimeter. Then, we represent the true location of the $j$-th UAV$_{A}$ in time slot $t$ as ${{\bf{q}}_{j,t}} \in {\mathbb{R}^{2 \times 1}}$ (${\bf{q}}_{j,t}^{3D} \!=\! {\left[ {{{\left( {{{\bf{q}}_{j,t}}} \right)}^T},{h_v}} \right]^T} \!\in {\mathbb{R}^{3 \times 1}}$). UAV$_{B}$ fly at the same altitude (${h_v}$) as UAV$_{A}$, and the location of the $i$-th UAV$_{B}$ ($i \in {{\cal N}_B}$) is denoted by ${{\bf{b}}_i} \in {\mathbb{R}^{2 \times 1}}$ (${\bf{b}}_i^{3D} \!=\! {\left[ {{{\left( {{{\bf{b}}_i}} \right)}^T},{h_v}} \right]^T} \!\in {\mathbb{R}^{3 \times 1}}$). UEs on the ground are represented by the set ${\cal M} = \left\{ {1, \cdots ,M} \right\}$, and the true location of the $m$-th UE ($m \in {\cal M}$) is denoted as ${{\bf{p}}_m} \in {\mathbb{R}^{2 \times 1}}$ (${\bf{p}}_m^{3D} \!=\! {\left[ {{{\left( {{{\bf{p}}_m}} \right)}^T},0} \right]^T} \!\in {\mathbb{R}^{3 \times 1}}$).
\begin{figure}[!t]
\centering
\includegraphics[height=1.9in,width=3.45in]{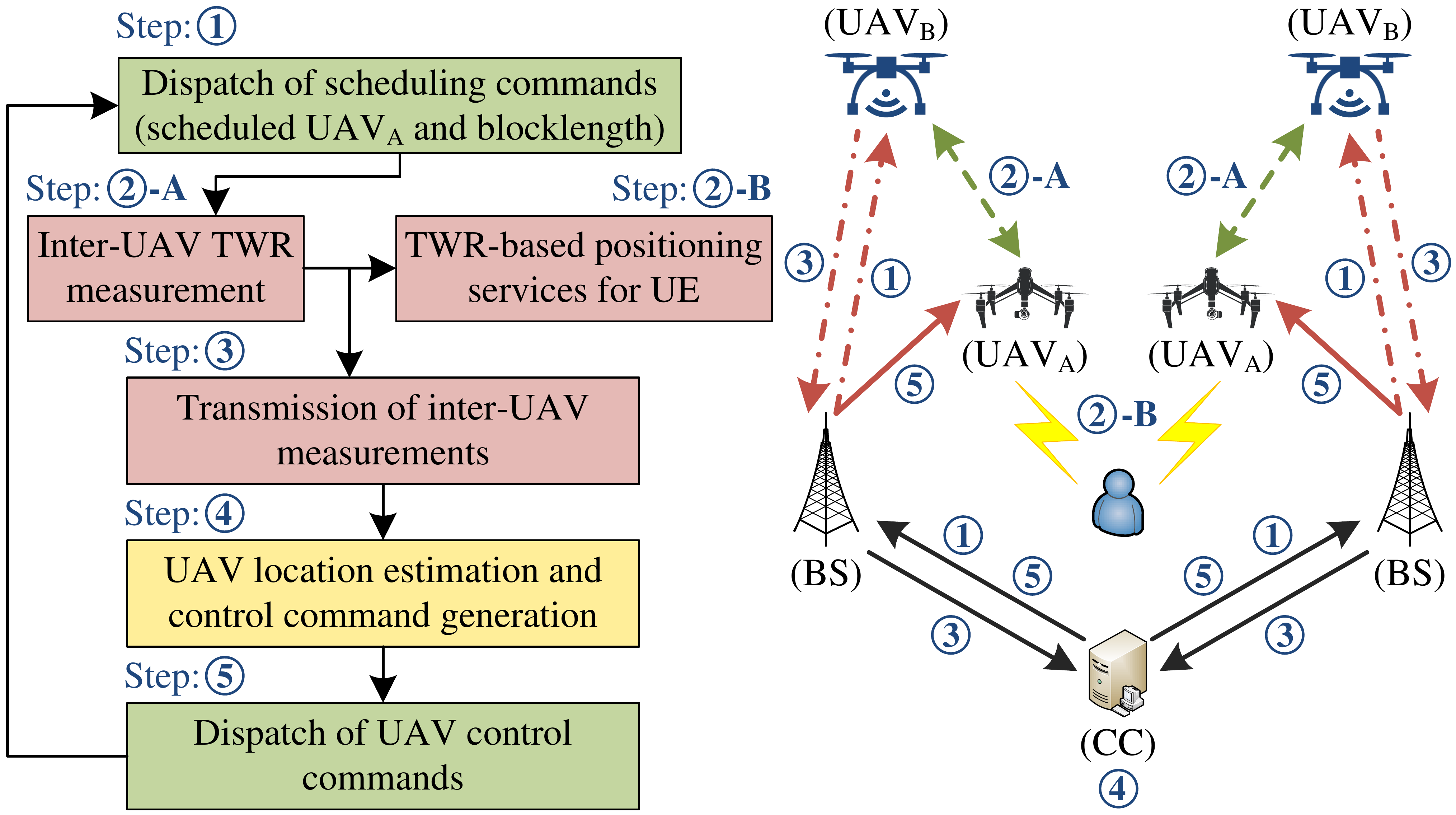}
\caption{Structure and operation strategy of the proposed system.}
\label{fig_2}
\end{figure}

Fig. 2 shows the structure and operation strategy of the proposed system. At the beginning of each time slot, the control center (CC) sends scheduling commands to all UAV$_{A}$ through terrestrial BSs and ground-to-air (G2A) communication links. The scheduling commands contain the index of the UAV$_{A}$ that need to perform state sensing in this time slot and the blocklength for the transmission of sensing data. Since terrestrial BSs commonly have large transmit power, the dispatch of scheduling commands is assumed to be perfect without any failure. After receiving the scheduling commands, the scheduled UAV$_{A}$ cooperate with UAV$_{B}$ to measure the corresponding inter-UAV distances, and send the velocity and acceleration measurements obtained by onboard sensors to UAV$_{B}$. Subsequently, UAV$_{A}$ provide positioning services to UEs utilizing the two-way ranging (TWR) technique, while UAV$_{B}$ transmit the sensing data obtained in the previous step to the terrestrial BSs and CC. The A2G channels between UAV$_{B}$ and terrestrial BSs are modeled as Rayleigh fading channels. Due to the time-varying nature of fading channel and UAVs' limited transmit power, failures may occur in the transmission of sensing data, which affect the state estimation and control of UAV$_{A}$. If the data required for the state sensing of a UAV$_{A}$ is successfully received by BSs, CC will estimate its current state and generate the control input command based on closed-loop (CL) operation mode. Otherwise, the UAV$_{A}$ will be controlled in open-loop (OP) mode based on its previously estimated states. At the end of this time slot, CC sends control commands to UAV$_{A}$, and each UAV$_{A}$ adjusts its state according to the received command. Similar to scheduling commands, the dispatch of control commands is also assumed to be perfect.

With the structure and operation strategy introduced above, the proposed system is expected to work efficiently and provide positioning services that meet users' requirements. In the following subsections, we provide the key mathematical models used in our system.

\subsection{Model of UAV State Sensing}
In the proposed system, the true state of the $j$-th UAV$_{A}$ in time slot $t$ can be represented by the following vector:
\begin{equation}
{{\bf{x}}_{j,t}} = {\left[ {\Delta {\bf{q}}_{j,t}^T,{\bf{v}}_{j,t}^T,{\bf{a}}_{j,t}^T} \right]^T} \in {\mathbb{R}^{6 \times 1}},
\end{equation}
where $\Delta {{\bf{q}}_{j,t}} \!=\! {\left[ {\Delta q_{j,t}^{\left( x \right)},\Delta q_{j,t}^{\left( y \right)}} \right]^T} \!=\! {{\bf{q}}_{j,t}} \!-\! {\bf{q}}_j^ \circ $ denotes the deviation between the $j$-th UAV$_{A}$'s true location and its corresponding HP; ${{\bf{v}}_{j,t}} \!=\! {\left[ {v_{j,t}^{\left( x \right)},v_{j,t}^{\left( y \right)}} \right]^T} \!\in\! {\mathbb{R}^{2 \times 1}}$ and ${{\bf{a}}_{j,t}} \!=\! {\left[ {a_{j,t}^{\left( x \right)},a_{j,t}^{\left( y \right)}} \right]^T} \!\in\! {\mathbb{R}^{2 \times 1}}$ are the true values of the UAV$_{A}$'s velocity and acceleration, respectively. Under ideal conditions without sensing and control errors, the UAV$_{A}$ should hover stably at the specified HP, i.e., ${{\bf{x}}_{j,t}} = {{\bf{0}}_{6 \times 1}}$.

In order to estimate the UAV$_{A}$'s location, the well-known TWR technique is used to measure the inter-UAV distances between UAV$_{A}$ and UAV$_{B}$. As derived in Appendix A, the relative range measurement corresponding to the $j$-th UAV$_{A}$ and $i$-th UAV$_{B}$ can be expressed as
\begin{equation}
\hat d_{j,t}^i = d_{j,t}^i + n_{d,j,t}^i = {\left\| {{\bf{q}}_{j,t}^{3D} - {\bf{b}}_i^{3D}} \right\|_2} + n_{d,j,t}^i,
\end{equation}
where $d_{j,t}^i = {\left\| {{\bf{q}}_{j,t}^{3D} - {\bf{b}}_i^{3D}} \right\|_2}$ denotes the true distance between the $j$-th UAV$_{A}$ and $i$-th UAV$_{B}$; $n_{d,j,t}^i \sim {\cal N}\left( {0,\sigma _d^2} \right)$ is the distance measurement error caused by the clock drifts of UAVs' local clocks, and $\sigma _d^2$ is its variance. As mentioned in \cite{UAV_Geo}, the value of $\sigma _d^2$ is mainly determined by the crystal tolerance of UAV's oscillator and response delay of TWR. Since UAVs in the proposed system use the same type of oscillators and the same TWR protocol, all inter-UAV distance measurements have the same variance ($\sigma _d^2$).

For each UAV$_{A}$ $j$, three UAV$_{B}$ are assigned to sense its state, denoted by the set ${\cal N}_B^{{A_j}}$ (${\cal N}_B^{{A_j}} \in {{\cal N}_B}$). Then, the three inter-UAV distance measurements corresponding to the $j$-th UAV$_{A}$ can be represented by the following vector:
\begin{equation}
{{\bf{\hat d}}_{j,t}} = {\left[ { \cdots ,\hat d_{j,t}^i, \cdots } \right]^T} = {{\bf{d}}_{j,t}} + {{\bf{n}}_{{\bf{d}},j,t}},\quad i \in {\cal N}_B^{{A_j}},
\end{equation}
where ${{\bf{d}}_{j,t}} = {\left[ { \cdots ,d_{j,t}^i, \cdots } \right]^T}$ ($i \in {\cal N}_B^{{A_j}}$); ${{\bf{n}}_{{\bf{d}},j,t}} = {\left[ { \cdots ,n_{d,j,t}^i, \cdots } \right]^T}$ is the noise vector consisting of mutually independent measurement errors with same variance. Thus, the covariance matrix of ${{\bf{n}}_{{\bf{d}},j,t}}$ can be expressed as ${{\bf{R}}_{\bf{d}}} = \sigma _d^2 \cdot {{\bf{I}}_3}$.

With the above measurement equations, the $j$-th UAV$_{A}$'s location in time slot $t$ can be estimated by CC through the maximum-likelihood (ML) method. Then, the Cram\'er-Rao lower bound (CRLB) that can be approached by the ML method is used to indicate the accuracy of the estimation of UAV$_{A}$'s locations, which can be expressed as
\begin{equation}
{\mathop{\rm CRLB}\nolimits} \left( {{{\bf{q}}_{j,t}}} \right) = {\left( {{\bf{H}}_j^T{\bf{R}}_{\bf{d}}^{ - 1}{{\bf{H}}_j}} \right)^{ - 1}} = \sigma _d^2 \cdot {\left( {{\bf{H}}_j^T{{\bf{H}}_j}} \right)^{ - 1}},
\end{equation}
where ${{\bf{H}}_j}$ is the Jacobian matrix of equation (3) at ${{\bf{q}}_{j,t}}$, and
\begin{equation}
{{\bf{H}}_j} = {{\partial {{\bf{d}}_{j,t}}} \mathord{\left/
 {\vphantom {{\partial {{\bf{d}}_{j,t}}} {\partial {{\bf{q}}_{j,t}}}}} \right.
 \kern-\nulldelimiterspace} {\partial {{\bf{q}}_{j,t}}}} = {\left[ { \cdots\! ,{\bf{h}}_j^i\left( {{{\bf{q}}_{j,t}}} \right), \!\cdots } \right]^T} = {\left[ { \cdots \!,{{{{\left( {{{\bf{q}}_{j,t}} \!-\! {{\bf{b}}_i}} \right)} \mathord{\left/
 {\vphantom {{\left( {{{\bf{q}}_{j,t}} - {{\bf{b}}_i}} \right)} {\left\| {{\bf{q}}_{j,t}^{3D} \!-\! {\bf{b}}_i^{3D}} \right\|}}} \right.
 \kern-\nulldelimiterspace} {\left\| {{\bf{q}}_{j,t}^{3D} - {\bf{b}}_i^{3D}} \right\|}}}_2}, \!\cdots } \right]^T},\;i \in {\cal N}_B^{{A_j}}.
\end{equation}

It is noteworthy that with appropriate control strategy, the deviation between UAV$_{A}$'s true location (${{\bf{q}}_{j,t}}$) and the corresponding HP (${\bf{q}}_j^ \circ $) is acceptable, and its influence on the geometry of UAVs is negligible \cite{UAV_Unc_1,Geo_Neg}. Thus, by replacing ${{\bf{q}}_{j,t}}$ in equation (5) with ${\bf{q}}_j^ \circ $, ${{\bf{H}}_j}$ can be approximated as
\begin{equation}
{{\bf{H}}_j} \approx {\left[ { \cdots\! ,{\bf{h}}_j^i\left( {{\bf{q}}_j^ \circ } \right), \!\cdots } \right]^T} = {\left[ { \cdots\! ,{{\left( {{\bf{q}}_j^ \circ  - {{\bf{b}}_i}} \right)} \mathord{\left/
 {\vphantom {{\left( {{\bf{q}}_j^ \circ  - {{\bf{b}}_i}} \right)} {{{\left\| {{\bf{q}}_j^{ \circ ,3D} \!-\! {\bf{b}}_i^{3D}} \right\|}_2}}}} \right.
 \kern-\nulldelimiterspace} {{{\left\| {{\bf{q}}_j^{ \circ ,3D} \!-\! {\bf{b}}_i^{3D}} \right\|}_2}}}, \!\cdots } \right]^T}.
\end{equation}

We assume that the estimation of UAV$_{A}$'s location performed at CC could approach the CRLB. Then, the covariance matrix of UAV$_{A}$ location estimate (${{\bf{\hat q}}_{j,t}}$) can be expressed as ${{\bf{R}}_{{\bf{q}},j}} \approx {\mathop{\rm CRLB}\nolimits} \left( {{{\bf{q}}_{j,t}}} \right)$. Moreover, utilizing the relationship $\Delta {{\bf{q}}_{j,t}} = {{\bf{q}}_{j,t}} - {\bf{q}}_j^ \circ $, the estimate of $\Delta {{\bf{q}}_{j,t}}$ can be written as
\begin{equation}
\Delta {{\bf{\hat q}}_{j,t}} = {{\bf{\hat q}}_{j,t}} - {\bf{q}}_j^ \circ  = \Delta {{\bf{q}}_{j,t}} + {{\bf{n}}_{\Delta {\bf{q}},j,t}},
\end{equation}
where ${{\bf{n}}_{\Delta {\bf{q}},j,t}} \sim {\cal N}\left( {{{\bf{0}}_{2 \times 1}},{{\bf{R}}_{\Delta {\bf{q}},j}}} \right)$ is the location estimation error, and ${{\bf{R}}_{\Delta {\bf{q}},j}} = {{\bf{R}}_{{\bf{q}},j}}$ is its covariance matrix.

In addition, while measuring the inter-UAV distances, UAV$_{A}$ also send their velocity and acceleration measurements to UAV$_{B}$ by embedding them into the response message of TWR. The velocity and acceleration of the $j$-th UAV$_{A}$ measured in time slot $t$ can be expressed as
\begin{equation}
{{\bf{\hat v}}_{j,t}} = {{\bf{v}}_{j,t}} + {{\bf{n}}_{{\bf{v}},j,t}},\qquad\qquad{{\bf{\hat a}}_{j,t}} = {{\bf{a}}_{j,t}} + {{\bf{n}}_{{\bf{a}},j,t}},
\end{equation}
where ${{\bf{n}}_{{\bf{v}},j,t}} \sim {\cal N}\left( {{{\bf{0}}_{2 \times 1}},{{\bf{R}}_{\bf{v}}}} \right)$ and ${{\bf{n}}_{{\bf{a}},j,t}} \sim {\cal N}\left( {{{\bf{0}}_{2 \times 1}},{{\bf{R}}_{\bf{a}}}} \right)$ represent measurement errors. Since we assume that all UAV$_{A}$ are equipped with the same sensors, the velocity (acceleration) measurements in the proposed system have the same covariance matrix ${{\bf{R}}_{\bf{v}}}$ (${{\bf{R}}_{\bf{a}}}$).

Then, the sensing results of the $j$-th UAV$_{A}$'s state in time slot $t$ can be represented by the following vector:
\begin{equation}
{{\bf{\hat x}}_{j,t}} = {\left[ {\Delta {\bf{\hat q}}_{j,t}^T,{\bf{\hat v}}_{j,t}^T,{\bf{\hat a}}_{j,t}^T} \right]^T} = {{\bf{x}}_{j,t}} + {{\bm{\eta }}_{j,t}},
\end{equation}
where ${{\bm{\eta }}_{j,t}} = {\left[ {{{\bf{n}}_{\Delta {\bf{q}},j,t}},{{\bf{n}}_{{\bf{v}},j,t}},{{\bf{n}}_{{\bf{a}},j,t}}} \right]^T} \sim {\cal N}\left( {{{\bf{0}}_{6 \times 1}},{{\bf{R}}_{{\bm{\eta }},j}}} \right)$ is the noise vector consisting of three kinds of mutually independent measurement errors, and its covariance matrix can be written as
\begin{equation}
{{\bf{R}}_{\bm{\eta },j}} = {\rm blkdiag}\left( {{{\bf{R}}_{\Delta {\bf{q}},j}},{{\bf{R}}_{\bf{v}}},{{\bf{R}}_{\bf{a}}}} \right).
\end{equation}

\subsection{Model of UAV Control}
For simplicity and without loss of generality, we model each UAV$_{A}$ in the proposed system as a linear time-invariant discrete-time system, whose dynamics model is given by
\begin{equation}
{{\bf{x}}_{j,t + 1}} = {{\bf{A}}_j}{{\bf{x}}_{j,t}} + {{\bf{B}}_j}{{\bf{u}}_{j,t}} + {{\bf{w}}_{j,t}},
\end{equation}
where ${{\bf{A}}_j} \in {\mathbb{R}^{6 \times 6}}$ and ${{\bf{B}}_j} \in {\mathbb{R}^{6 \times 2}}$ are the state matrix and input matrix of the $j$-th UAV$_{A}$, respectively; ${{\bf{u}}_{j,t}} = {\left[ {u_j^{\left( x \right)},u_j^{\left( y \right)}} \right]^T} \in {\mathbb{R}^{2 \times 1}}$ is the control input, i.e., the commanded acceleration; ${{\bf{w}}_{j,t}} \sim {\cal N}\left( {{{\bf{0}}_{6 \times 1}},{{\bf{Q}}_{\bf{w}}}} \right)$ is independent identically distributed (i.i.d.) process noise that characterizes the unintentional movements of UAV$_{A}$. The expressions for matrices ${{\bf{A}}_j}$, ${{\bf{B}}_j}$ and ${{\bf{Q}}_{\bf{w}}}$ are derived in Appendix B.

As a linear system, UAV$_{A}$'s control input vector can be determined using the state feedback law ${{\bf{u}}_{j,t}} = {{\bf{K}}_j}{{\bf{x}}_{j,t}}$, where ${{\bf{K}}_j} \in {\mathbb{R}^{2 \times 6}}$ is the gain matrix. Then, equation (11) can be rewritten as
\begin{equation}
{{\bf{x}}_{j,t + 1}} = \left( {{{\bf{A}}_j} + {{\bf{B}}_j}{{\bf{K}}_j}} \right){{\bf{x}}_{j,t}} + {{\bf{w}}_{j,t}}.
\end{equation}

Please note that the true state (${{\bf{x}}_{j,t}}$) in the above equation is unavailable in practice. Therefore, CC can only generate the control input (${{\bf{u}}_{j,t}}$) based on state sensing or prediction results. If the sensing data is successfully received by CC, the control input will be generated in CL mode based on the estimated state (${{\bf{\hat x}}_{j,t}}$), i.e., ${\bf{u}}_{j,t}^{\left( c \right)} = {{\bf{K}}_j}{{\bf{\hat x}}_{j,t}} = {{\bf{K}}_j}\left( {{{\bf{x}}_{j,t}} + {{\bm{\eta }}_{j,t}}} \right)$. Otherwise, the UAV$_{A}$ will be controlled in OL mode and the control input is generated based on the predicted state. According to equation (12), if the last successful state sensing was performed in time slot $t - \Delta t_j^c$, the prediction of UAV$_{A}$'s state in time slot $t$ can be expressed as
\begin{equation}
{\bf{\bar x}}_{j,t}^{\left( {\Delta t_j^c} \right)} = {\left( {{{\bf{A}}_j} + {{\bf{B}}_j}{{\bf{K}}_j}} \right)^{\Delta t_j^c}}{{{\bf{\hat x}}}_{j,t - \Delta t_j^c}} = {\left( {{{\bf{A}}_j} \!+\! {{\bf{B}}_j}{{\bf{K}}_j}} \right)^{\Delta t_j^c}}\left( {{{\bf{x}}_{j,t - \Delta t_j^c}} \!+\! {{\bm{\eta }}_{j,t - \Delta t_j^c}}} \right).
\end{equation}
Then, the control input generated in OL mode can be written as ${\bf{u}}_{j,t}^{\left( o \right)} = {{\bf{K}}_j}{\bf{\bar x}}_{j,t}^{\left( {\Delta t_j^c} \right)}$, and the UAV$_{A}$'s dynamics model in two operation modes can be expressed as
\begin{equation}
\begin{split}
{\bf{x}}_{j,t + 1}^{\left( c \right)} &= \left( {{{\bf{A}}_j} + {{\bf{B}}_j}{{\bf{K}}_j}} \right){{\bf{x}}_{j,t}} + \left( {{{\bf{B}}_j}{{\bf{K}}_j}{{\bm{\eta }}_{j,t}} + {{\bf{w}}_{j,t}}} \right),\quad{\rm closed\raisebox{0mm}{-}loop},\\
{\bf{x}}_{j,t + 1}^{\left( o \right)} &= {{\bf{A}}_j}{{\bf{x}}_{j,t}} + {{\bf{B}}_j}{{\bf{K}}_j}{\bf{\bar x}}_{j,t}^{\left( {\Delta t_j^c} \right)} + {{\bf{w}}_{j,t}},\qquad\qquad\quad\;{\rm open\raisebox{0mm}{-}loop},
\end{split}
\end{equation}
where superscripts $\left( c \right)$ and $\left( o \right)$ indicate the CL and OL modes, respectively.

Furthermore, the update rule for parameter $\Delta t_j^c$ in two operation modes is given by
\begin{equation}
\Delta t_j^c \leftarrow \left\{ {\begin{array}{*{20}{c}}
{1,\qquad\qquad{\rm closed\raisebox{0mm}{-}loop},}\\
{\Delta t_j^c + 1,\quad\;\,{\rm open\raisebox{0mm}{-}loop}.\;\;}
\end{array}} \right.
\end{equation}

\subsection{Model of Positioning Services for UE}
As mentioned at the beginning of this section, in the proposed system, UAV$_{A}$ are employed as aerial anchor nodes to provide TWR-based positioning services for UEs. Similar to the measurement of inter-UAV distances, the distance measurement between the $m$-th UE and $j$-th UAV$_{A}$ obtained through the TWR technique in time slot $t$ can be expressed as
\begin{equation}
\hat r_{m,t}^j = r_{m,t}^j + n_{r,m,t}^j = {\left\| {{\bf{p}}_m^{3D} - {\bf{q}}_{j,t}^{3D}} \right\|_2} + n_{r,m,t}^j,
\end{equation}
where $r_{m,t}^j = {\left\| {{\bf{p}}_m^{3D} - {\bf{q}}_{j,t}^{3D}} \right\|_2}$ denotes the true distance; $n_{r,m,t}^j \sim {\cal N}\left( {0,\sigma _r^2} \right)$ is the i.i.d. measurement error, and $\sigma _r^2$ is its variance.

Denote the $N_A^U$ UAV$_{A}$ ($N_A^U \ge 3$) serving the $m$-th UE as the set ${\cal N}_A^{{U_m}}$ (${\cal N}_A^{{U_m}} \in {{\cal N}_A}$), then the distance measurements available to the UE form the following vector:
\begin{equation}
{{\bf{\hat r}}_{m,t}} = {\left[ { \cdots ,\hat r_{m,t}^j, \cdots } \right]^T} = {{\bf{r}}_{m,t}} + {{\bf{n}}_{{\bf{r}},m,t}},\quad j \in {\cal N}_A^{{U_m}},
\end{equation}
where ${{\bf{r}}_{m,t}} = {\left[ { \cdots ,r_{m,t}^j, \cdots } \right]^T}$ ($j \in {\cal N}_A^{{U_m}}$); ${{\bf{n}}_{{\bf{r}},m,t}} = {\left[ { \cdots ,n_{r,m,t}^j, \cdots } \right]^T}$ is the noise vector consisting of $N_A^U$ i.i.d. measurement errors, and its covariance matrix can be written as ${{\bf{R}}_{\bf{r}}} = \sigma _r^2 \cdot {{\bf{I}}_{N_A^U}}$.

Due to unknown sensing and control errors, the true locations of UAV$_{A}$ are time-varying and unavailable to UEs. Thus, UEs can only use UAV$_{A}$'s HPs and the following measurement equations for position estimation:
\begin{equation}
{{{\bf{\bar r}}}_m}\left( {{{\bf{p}}_m}} \right) = {\left[ { \cdots ,\bar r_m^j\left( {{{\bf{p}}_m}} \right), \cdots } \right]^T} = {\left[ { \cdots ,{{\left\| {{\bf{p}}_m^{3D} - {\bf{q}}_j^{ \circ ,3D}} \right\|}_2}, \cdots } \right]^T}.
\end{equation}

In this article, the iterative least-squares (ILS) method that estimates unknown parameters in an iterative manner through Taylor-series linearization is used to determine UEs' locations \cite{Pos_ILS}. Denote the $m$-th UE's location estimate obtained in the $l$-th iteration of ILS as ${\bf{\hat p}}_{m,t}^{\left( l \right)} \in {\mathbb{R}^{2 \times 1}}$ (${\bf{\hat p}}_{m,t}^{\left( l \right),3D} = {\left[ {{{\left( {{\bf{\hat p}}_{m,t}^{\left( l \right)}} \right)}^T},0} \right]^T}$), then the first-order Taylor-series expansion of ${{\bf{\bar r}}_m}\left( {{{\bf{p}}_m}} \right)$ at ${\bf{\hat p}}_{m,t}^{\left( l \right)}$ can be written as
\begin{equation}
{{\bf{\bar r}}_m}\left( {{{\bf{p}}_m}} \right) \simeq {{\bf{\bar r}}_m}\left( {{\bf{\hat p}}_{m,t}^{\left( l \right)}} \right) + {{\bf{H}}_m}\left( {{\bf{\hat p}}_{m,t}^{\left( l \right)}} \right)\left( {{{\bf{p}}_m} - {\bf{\hat p}}_{m,t}^{\left( l \right)}} \right),
\end{equation}
where
\begin{equation}
{{\bf{H}}_m}\left(\! {{\bf{\hat p}}_{m,t}^{\left( l \right)}} \!\right) = {\left. {\frac{{\partial {{{\bf{\bar r}}}_m}\left( {{{\bf{p}}_m}} \right)}}{{\partial {{\bf{p}}_m}}}} \right|_{{\bf{\hat p}}_{m,t}^{\left( l \right)}}} \!=\! {\left[ { \cdots ,{\bf{h}}_m^j\left(\! {{\bf{\hat p}}_{m,t}^{\left( l \right)}} \!\right), \cdots } \right]^T} = {\left[ { \cdots ,\frac{{\left( {{\bf{\hat p}}_{m,t}^{\left( l \right)} \!-\! {\bf{q}}_j^ \circ } \right)}}{{{{\left\| {{\bf{\hat p}}_{m,t}^{\left( l \right),3D} \!-\! {\bf{q}}_j^{ \circ ,3D}} \right\|}_2}}}, \cdots } \right]^T}
\end{equation}
is the Jacobian matrix of equation (18) at ${\bf{\hat p}}_{m,t}^{\left( l \right)}$.

The LS estimate of the $m$-th UE's location obtained in the $\left( {l + 1} \right)$-th iteration is given by
\begin{equation}
{\bf{\hat p}}_{m,t}^{\left( {l + 1} \right)} = {\bf{\hat p}}_{m,t}^{\left( l \right)} + {\bf{S}}\left( {{\bf{\hat p}}_{m,t}^{\left( l \right)}} \right)\left[ {{{{\bf{\hat r}}}_{m,t}} - {{{\bf{\bar r}}}_m}\left( {{\bf{\hat p}}_{m,t}^{\left( l \right)}} \right)} \right],
\end{equation}
where
\begin{equation}
{\bf{S}}\left( {{\bf{\hat p}}_{m,t}^{\left( l \right)}} \right) = {\bf{P}}\left( {{\bf{\hat p}}_{m,t}^{\left( l \right)}} \right){\bf{H}}{\left( {{\bf{\hat p}}_{m,t}^{\left( l \right)}} \right)^T},
\end{equation}
\begin{equation}
{\bf{P}}\left( {{\bf{\hat p}}_{m,t}^{\left( l \right)}} \right) = {\left[ {{\bf{H}}{{\left( {{\bf{\hat p}}_{m,t}^{\left( l \right)}} \right)}^T}{\bf{H}}\left( {{\bf{\hat p}}_{m,t}^{\left( l \right)}} \right)} \right]^{ - 1}}.
\end{equation}

The estimation result of the $m$-th UE's location in time slot $t$ can be obtained after the above iteration converges.

\section{Performance Analysis and Problem Formulation}
In this article, the UAV position uncertainty caused by sensing and control errors is taken into account in the design of our UAV-enabled positioning system, which has often been overlooked in existing research. Due to this additional consideration, the analysis methods and performance metrics used in previous research are unsuitable for evaluating the performance of the proposed system. Therefore, in the first two subsections of this section, we first analysis the stability of UAV$_{A}$ in two operation modes, and then derive the MSE of UE positioning in the presence of UAV position uncertainty. It is found that the scheduling of UAV$_{A}$ state sensing and the blocklength for the transmission of sensing data are two main factors affecting the quality of positioning services. Thus, at the end of this section, we formulate the sensing scheduling and blocklength allocation in the proposed system as a QoS-constrained optimization problem, which will be solved using the scheme introduced in the next section.

\subsection{UAV Stability Prediction in Different Operation Modes}
The aim of this subsection is to predict the stability of each UAV$_{A}$ in the next time slot ($t + 1$) before CC generates the scheduling commands for the current time slot ($t$). The UAV$_{A}$'s dynamics model in two operation modes has been derived in Section II.B. However, it is noteworthy that the dynamics model shown in equation (14) uses the true state of UAV$_{A}$ (${{\bf{x}}_{j,t}}$) when calculating its future state, which is unknown in practice. Therefore, in order to predict UAV$_{A}$'s state in time slot $t + 1$, we need to replace ${{\bf{x}}_{j,t}}$ in equation (14) with other variables whose values or distributions are available to CC. We assume that the last successful state sensing was performed $t + 1$ time slots ago. Then, according to equations (13) and (14), the expression for ${{\bf{x}}_{j,t}}$ can be written as
\begin{equation}
{{\bf{x}}_{j,t}} \!=\! {\left(\! {{{\bf{A}}\!_j} \!+\! {{\bf{B}}\!_j}{{\bf{K}}\!_j}} \!\right)^{\Delta t_j^c}}{{\bf{x}}_{j,t - \Delta t_j^c}} + \left[\! {\sum\limits_{k = 0}^{\Delta t_j^c \!-\! 1}\! {{\bf{A}}\!_j^k{{\bf{B}}\!_j}{{\bf{K}}\!_j}{{\left(\! {{{\bf{A}}\!_j} \!+\! {{\bf{B}}\!_j}{{\bf{K}}\!_j}} \!\right)}^{\Delta t_j^c \!-\! k \!-\! 1}}} } \!\right]\!{{\bm{\eta }}_{j,t \!-\! \Delta t_j^c}} \!+\! \sum\limits_{k = 0}^{\Delta t_j^c - 1}\! {{\bf{A}}\!_j^k{{\bf{w}}_{j,t - k - 1}}} .
\end{equation}
Moreover, utilizing the relationship ${{\bf{x}}_{j,t - \Delta t_j^c}} = {{\bf{\hat x}}_{j,t - \Delta t_j^c}} - {{\bm{\eta }}_{j,t - \Delta t_j^c}}$ and equation (13), the above equation can be further rewritten as
\begin{equation}
{{\bf{x}}_{j,t}} \!=\! {\bf{\bar x}}_{j,t}^{\left(\! {\Delta t_j^c} \!\right)} \!+\! \left[\! {\left(\! {\sum\limits_{k = 0}^{\Delta t_j^c \!-\! 1}\! {{\bf{A}}\!_j^k{{\bf{B}}_j}{{\bf{K}}_j}{{\left(\! {{{\bf{A}}\!_j} \!+\! {{\bf{B}}_j}{{\bf{K}}_j}} \!\right)}\!^{\Delta t_j^c \!-\! k \!-\! 1}}} } \!\right) \!-\! {{\left(\! {{{\bf{A}}\!_j} \!+\! {{\bf{B}}_j}{{\bf{K}}_j}} \!\right)}\!^{\Delta t_j^c}}} \!\right]\!{{\bm{\eta }}_{j,t \!-\! \Delta t_j^c}} \!+\! \sum\limits_{k = 0}^{\Delta t_j^c \!-\! 1} \!{{\bf{A}}\!_j^k{{\bf{w}}_{j,t \!-\! k \!-\! 1}}} .
\end{equation}
The value of ${\bf{\bar x}}_{j,t}^{\left( {\Delta t_j^c} \right)}$ in the above equation can be calculated using equation (13), and the distributions of noise vectors ${{\bm{\eta }}_{j,t - \Delta t_j^c}}$ and ${{\bf{w}}_{j,t - k - 1}}$ are known. Then, by substituting this equation into equation (14), the UAV$_{A}$'s dynamics model in two operation modes can be rewritten as
\begin{equation}
\begin{split}
{\bf{x}}\!_{j,t \!+\! 1}^{\left( c \right)} &\!=\! {\bf{A}}\!_j^{\left(\! c \!\right)}{\bf{\bar x}}\!_{j,t}^{\left(\! {\Delta t_j^c} \!\right)} \!+\! \left(\! {{\bf{A}}\!_j^{\left(\! c \!\right)}\!{\bf{C}}\!_j^{\left(\! {\Delta t_j^c} \!\right)}\!{{\bm{\eta }}_{j,t \!-\! \Delta t_j^c}} \!+\! {{\bf{B}}\!_j}{{\bf{K}}\!_j}{{\bm{\eta }}_{j,t}}} \!\right) \!+\! \left(\! {\sum\limits_{k = 0}^{\Delta t_j^c \!-\! 1}\! {{\bf{A}}\!_j^{\left(\! c \!\right)}\!{\bf{A}}\!_j^k{{\bf{w}}\!_{j,t \!-\! k \!-\! 1}}}  \!+\! {{\bf{w}}\!_{j,t}}} \!\right)\!,{\rm closed\raisebox{0mm}{-}loop},\\
{\bf{x}}_{j,t \!+\! 1}^{\left( o \right)} &\!=\! {\bf{A}}\!_j^{\left(\! c \!\right)}{\bf{\bar x}}_{j,t}^{\left(\! {\Delta t_j^c} \!\right)} \!+\! {{\bf{A}}\!_j}{\bf{C}}\!_j^{\left(\! {\Delta t_j^c} \!\right)}{{\bm{\eta }}_{j,t \!-\! \Delta t_j^c}} \!+\! \sum\limits_{k = 0}^{\Delta t_j^c}\! {{\bf{A}}\!_j^k{{\bf{w}}\!_{j,t \!-\! k}}} ,\qquad\qquad\qquad\qquad\qquad\quad\;{\rm open\raisebox{0mm}{-}loop},
\end{split}
\end{equation}
where
\begin{equation}
{\bf{A}}_j^{\left( c \right)} = {{\bf{A}}_j} + {{\bf{B}}_j}{{\bf{K}}_j}
\end{equation}
is the closed loop state matrix, and
\begin{equation}
{\bf{C}}_j^{\left( {\Delta t_j^c} \right)} = \left[ {\sum\limits_{k = 0}^{\Delta t_j^c - 1} {{\bf{A}}_j^k{{\bf{B}}_j}{{\bf{K}}_j}{{\left( {{\bf{A}}_j^{\left( c \!\right)}} \right)}^{\Delta t_j^c - k - 1}}}  - {{\left(\! {{\bf{A}}_j^{\left( c \right)}} \right)}^{\Delta t_j^c}}} \right].
\end{equation}

Please note that the values of vectors ${{\bm{\eta }}_{j,t - \Delta t_j^c}}$ and ${{\bf{w}}_{j,t - k - 1}}$ in equation (26) are unavailable. Therefore, we choose to predict the covariance matrix of UAV$_{A}$'s state in time slot $t$ instead of its true value. According to equation (26), the predicted covariance matrix in two operation modes can be expressed as
\begin{equation}
\begin{split}
{\bf{Q}}_{{\bf{x}},j,t \!+\! 1}^{\left( c \right)} &\!=\! E\!\left\{\! {\left(\! {{\bf{x}}\!_{j,t \!+\! 1}^{\left( c \right)}} \!\right)\!{{\left(\! {{\bf{x}}_{j,t \!+\! 1}^{\left( c \right)}} \!\right)}^T}} \!\right\} \!=\! \left(\! {{\bf{A}}\!_j^{\left(\! c \!\right)}{\bf{\bar x}}_{j,t}^{\left(\! {\Delta t_j^c} \!\right)}} \!\right)\!{\left(\! {{\bf{A}}\!_j^{\left(\! c \!\right)}{\bf{\bar x}}_{j,t}^{\left(\! {\Delta t_j^c} \!\right)}} \!\right)^T} \!+\! \left(\! {{\bf{A}}\!_j^{\left(\! c \!\right)}{\bf{C}}_j^{\left(\! {\Delta t_j^c} \!\right)}} \!\right)\!{{\bf{R}}_{\bm{\eta },j}}\!{\left(\! {{\bf{A}}\!_j^{\left(\! c \!\right)}{\bf{C}}_j^{\left(\! {\Delta t_j^c} \!\right)}} \!\right)^T}\\
&\qquad\qquad\qquad\qquad\quad\; + \left( {{{\bf{B}}_j}{{\bf{K}}_j}} \right)\!{{\bf{R}}_{\bm{\eta },j}}{\left( {{{\bf{B}}_j}{{\bf{K}}_j}} \right)^T} \!+\! \sum\limits_{k = 0}^{\Delta t_j^c \!-\! 1}\! {\left(\! {{\bf{A}}\!_j^{\left(\! c \!\right)}{\bf{A}}\!_j^k} \right)\!{{\bf{Q}}_{\bf{w}}}\!{{\left(\! {{\bf{A}}\!_j^{\left(\! c \!\right)}{\bf{A}}\!_j^k} \!\right)}^T}}  \!+\! {{\bf{Q}}_{\bf{w}}},\\
{\bf{Q}}_{{\bf{x}},j,t \!+\! 1}^{\left( o \right)} &\!=\! E\!\left\{\! {\left(\! {{\bf{x}}_{j,t \!+\! 1}^{\left( o \right)}} \!\right)\!{{\left(\! {{\bf{x}}_{j,t \!+\! 1}^{\left( o \right)}} \!\right)}^T}} \!\right\} \!=\! \left(\! {{\bf{A}}\!_j^{\left(\! c \!\right)}{\bf{\bar x}}_{j,t}^{\left(\! {\Delta t_j^c} \!\right)}} \!\right)\!{\left(\! {{\bf{A}}\!_j^{\left(\! c \!\right)}{\bf{\bar x}}_{j,t}^{\left(\! {\Delta t_j^c} \!\right)}} \!\right)^T} \!+\! \left(\! {{{\bf{A}}\!_j}{\bf{C}}_j^{\left(\! {\Delta t_j^c} \!\right)}} \!\right)\!{{\bf{R}}_{\bm{\eta },j}}\!{\left(\! {{{\bf{A}}\!_j}{\bf{C}}_j^{\left(\! {\Delta t_j^c} \!\right)}} \!\right)^T}\\
&\qquad\qquad\qquad\qquad\quad\; + \sum\limits_{k = 0}^{\Delta t_j^c}\! {\left( {{\bf{A}}\!_j^k} \right)\!{{\bf{Q}}_{\bf{w}}}\!{{\left( {{\bf{A}}\!_j^k} \right)}^T}.}
\end{split}
\end{equation}

In this article, we mainly focus on the UAV position uncertainty ($\Delta {{\bf{q}}_{j,t + 1}} = {\left( {{{\bf{x}}_{j,t + 1}}} \right)_{1:2}}$), whose covariance matrix can be written as
\begin{equation}
\begin{split}
{\bf{Q}}_{\Delta {\bf{q}},j,t + 1}^{\left( c \right)} &= {\left( {{\bf{Q}}_{{\bf{x}},j,t + 1}^{\left( c \right)}} \right)_{2 \times 2}},\quad\;{\rm closed\raisebox{0mm}{-}loop},\\
{\bf{Q}}_{\Delta {\bf{q}},j,t + 1}^{\left( o \right)} &= {\left( {{\bf{Q}}_{{\bf{x}},j,t + 1}^{\left( o \right)}} \right)_{2 \times 2}},\quad\;{\rm open\raisebox{0mm}{-}loop}.
\end{split}
\end{equation}

\subsection{UE Positioning Performance with UAV Control Error}
The covariance matrix of UAV position uncertainty has been derived in the previous subsection. Then, in this subsection, we further analyze the performance of positioning services in the presence of UAV control error. According to the model described in Section II.C, the estimation error of the $m$-th UE's location in the $\left( {l + 1} \right)$-th iteration of ILS can be expressed as
\begin{equation}
\Delta {\bf{p}}_{m,t + 1}^{\left( {l + 1} \right)} = {\bf{\hat p}}_{m,t + 1}^{\left( {l + 1} \right)} - {{\bf{p}}_m} = \left( {{\bf{\hat p}}_{m,t + 1}^{\left( l \right)} - {{\bf{p}}_m}} \right) + {\bf{S}}\left( {{\bf{\hat p}}_{m,t + 1}^{\left( l \right)}} \right)\left[ {{{{\bf{\hat r}}}_{m,t + 1}} - {{{\bf{\bar r}}}_m}\left( {{\bf{\hat p}}_{m,t + 1}^{\left( l \right)}} \right)} \right].
\end{equation}

With appropriate UAV control strategy and good initial guesses for ILS, the location estimate will be close to the true location after several iterations \cite{UAV_Unc_1}. Thus, by replacing ${\bf{\hat p}}_{m,t + 1}^{\left( l \right)}$ in the above equation with the true location (${{\bf{p}}_m}$), the estimation error of the ILS method after convergence can be written as
\begin{equation}
\Delta {{\bf{p}}_{m,t + 1}} = {{\bf{\hat p}}_{m,t + 1}} - {{\bf{p}}_m} = {\bf{S}}\left( {{{\bf{p}}_m}} \right)\left[ {{{{\bf{\hat r}}}_{m,t + 1}} - {{{\bf{\bar r}}}_m}\left( {{{\bf{p}}_m}} \right)} \right].
\end{equation}
The matrix ${\bf{S}}\left( {{{\bf{p}}_m}} \right)$ in the above equation can be easily calculated using equations (22) and (23). We then derive the expression for term $\left[ {{{{\bf{\hat r}}}_{m,t + 1}} - {{{\bf{\bar r}}}_m}\left( {{{\bf{p}}_m}} \right)} \right]$. Applying the first-order Taylor-series expansion to equation (16) based on the relationship ${{\bf{q}}_{j,t + 1}} = {\bf{q}}_j^ \circ  + \Delta {{\bf{q}}_{j,t + 1}}$ (${\bf{q}}_{j,t + 1}^{3D} = {\bf{q}}_j^{ \circ ,3D} + {\left[ {\Delta {\bf{q}}_{j,t + 1}^T,0} \right]^T}$), $\hat r_{m,t + 1}^j$ can be approximated as
\begin{equation}
\hat r_{m,t \!+\! 1}^j \!\simeq\! {\left\| {{\bf{p}}_m^{3D} \!-\! {\bf{q}}_j^{ \circ ,3D}} \right\|_2} \!- {\bf{h}}_m^j{\left( {{{\bf{p}}_m}} \right)^T}\Delta {{\bf{q}}_{j,t \!+\! 1}} + n_{r,m,t \!+\! 1}^j \!=\! \bar r_m^j\left( {{{\bf{p}}_m}} \right) - {\bf{h}}_m^j{\left( {{{\bf{p}}_m}} \right)^T}\Delta {{\bf{q}}_{j,t \!+\! 1}} + n_{r,m,t \!+\! 1}^j.
\end{equation}
Then, term $\left[ {{{{\bf{\hat r}}}_{m,t + 1}} - {{{\bf{\bar r}}}_m}\left( {{{\bf{p}}_m}} \right)} \right]$ can be rewritten as
\begin{equation}
{{{\bf{\hat r}}}_{m,t \!+\! 1}} - {{{\bf{\bar r}}}_m}\left( {{{\bf{p}}_m}} \right) =  -\! {\left[ { \cdots \!,{\bf{h}}_m^j{{\left( {{{\bf{p}}_m}} \right)}^T}\Delta {{\bf{q}}_{j,t \!+\! 1}}, \!\cdots } \right]^T} + {\left[ { \cdots \!,n_{r,m,t \!+\! 1}^j,\! \cdots } \right]^T} \!=\!  - {{\bm{\beta }}_{m,t \!+\! 1}} + {{\bf{n}}_{{\bf{r}},m,t \!+\! 1}}.
\end{equation}
where ${{\bm{\beta }}_{m,t + 1}} = {\left[ { \cdots ,{\bf{h}}_m^j{{\left( {{{\bf{p}}_m}} \right)}^T}\Delta {{\bf{q}}_{j,t + 1}}, \cdots } \right]^T}$.

Substituting the above equation into equation (32), the covariance matrix of the $m$-th UE's position error in time slot $t + 1$ can be expressed as
\begin{equation}
\begin{split}
{\bf{Q}}_{\Delta {\bf{p}},m,t \!+\! 1}^{\left( {{{\bm{\xi }}_{m,t}}} \right)} &=\! E\!\left\{\! {\Delta {{\bf{p}}_{m,t \!+\! 1}}\Delta {\bf{p}}_{m,t \!+\! 1}^T} \!\right\} \!=\! E\!\left\{\! {{\bf{S}}\!\left(\! {{{\bf{p}}_m}} \!\right)\left(\! { -\! {{\bm{\beta }}_{m,t \!+\! 1}} \!+\! {{\bf{n}}_{{\bf{r}},m,t \!+\! 1}}} \!\right)\!{{\left(\! { -\! {{\bm{\beta }}_{m,t \!+\! 1}} \!+\! {{\bf{n}}_{{\bf{r}},m,t \!+\! 1}}} \!\right)}^T}\!{\bf{S}}{{\left(\! {{{\bf{p}}_m}} \right)}^T}} \!\right\}\\
&=\! {\bf{S}}\!\left(\! {{{\bf{p}}_m}} \!\right)\left[ {E\left\{ {{{\bm{\beta }}_{m,t \!+\! 1}}{\bm{\beta }}_{m,t \!+\! 1}^T} \right\} + E\left\{ {{{\bf{n}}_{{\bf{r}},m,t \!+\! 1}}{\bf{n}}_{{\bf{r}},m,t \!+\! 1}^T} \right\}} \right]{\bf{S}}{\left(\! {{{\bf{p}}_m}} \!\right)^T}\\
&=\! \sigma _r^2{\bf{P}}\!\left(\! {{{\bf{p}}_m}} \!\right) + {\bf{S}}\!\left(\! {{{\bf{p}}_m}} \!\right){\mathop{\rm diag}\nolimits} \left( { \ldots ,{\bf{h}}_m^j{{\left( {{{\bf{p}}_m}} \right)}^T}{\bf{Q}}_{\Delta {\bf{q}},j,t \!+\! 1}^{\left( {{\xi _{m,j,t}}} \right)}{\bf{h}}_m^j\left( {{{\bf{p}}_m}} \right), \ldots } \right){\bf{S}}{\left(\! {{{\bf{p}}_m}} \!\right)^T},
\end{split}
\end{equation}
where vector ${{\bm{\xi }}_{m,t}} = {\left[ { \cdots ,{\xi _{m,j,t}}, \cdots } \right]^T}$ ($j \in {\cal N}_A^{{U_m}}$) represents the operation modes of the $N_A^U$ UAV$_{A}$ serving the $m$-th UE; ${\xi _{m,j,t}}$ could be either $c$ or $o$, indicating the operation mode of the $j$-th UAV$_{A}$ in time slot $t$; Covariance matrix ${\bf{Q}}_{\Delta {\bf{q}},j,t + 1}^{\left( {{\xi _{m,j,t}}} \right)}$ can be calculated using equations (29) and (30).
\begin{figure}[!t]
\centering
\includegraphics[height=1.65in,width=3.45in]{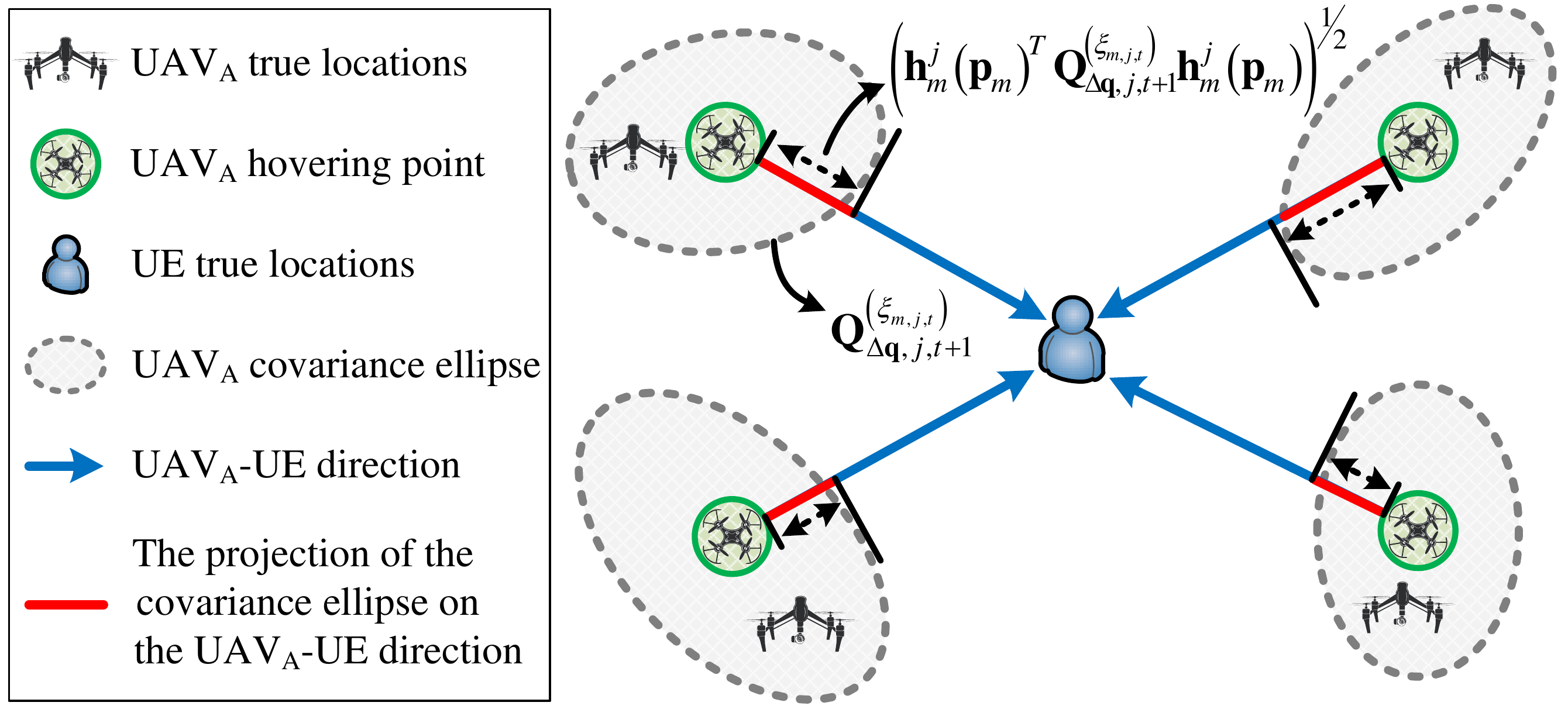}
\caption{Explanation of UAV control errors' impact on UE positioning.}
\label{fig_3}
\end{figure}

As can be seen from equation (35), the covariance matrix of UE position error can be written as the sum of two terms. The first term $\sigma _r^2{\bf{P}}\left( {{{\bf{p}}_m}} \right)$ denotes the position error caused by distance measurement errors, reflecting the positioning accuracy in the absence of UAV position uncertainty. The second term ${\bf{S}}\left( {{{\bf{p}}_m}} \right){\mathop{\rm diag}\nolimits} \left( { \ldots ,{\bf{h}}_m^j{{\left( {{{\bf{p}}_m}} \right)}^T}{\bf{Q}}_{\Delta {\bf{q}},j,t + 1}^{\left( {{\xi _{m,j,t}}} \right)}{\bf{h}}_m^j\left( {{{\bf{p}}_m}} \right), \ldots } \right){\bf{S}}{\left( {{{\bf{p}}_m}} \right)^T}$ indicates the performance degradation of positioning services caused by UAV control errors. Fig. 3 gives an intuitive explanation of UAV control errors' impact on UE positioning. According to equation (26), UAV control error can be represented as a linear combination of a set of independent Gaussian random variables. Thus, the control error also follows a Gaussian distribution and its covariance matrix (${\bf{Q}}_{\Delta {\bf{q}},j,t + 1}^{\left( {{\xi _{m,j,t}}} \right)}$) can be visually represented by the gray ellipse in Fig. 3. Then, the term ${\bf{h}}_m^j{\left( {{{\bf{p}}_m}} \right)^T}{\bf{Q}}_{\Delta {\bf{q}},j,t + 1}^{\left( {{\xi _{m,j,t}}} \right)}{\bf{h}}_m^j\left( {{{\bf{p}}_m}} \right)$ in equation (35) can be intuitively understood as the square of the projected length of the $j$-th UAV$_{A}$'s covariance ellipse in the UAV$_{A}$-UE direction. Finally, matrix ${\bf{S}}\left( {{{\bf{p}}_m}} \right)$ is used to convert term ${\bf{h}}_m^j{\left( {{{\bf{p}}_m}} \right)^T}{\bf{Q}}_{\Delta {\bf{q}},j,t + 1}^{\left( {{\xi _{m,j,t}}} \right)}{\bf{h}}_m^j\left( {{{\bf{p}}_m}} \right)$ into its contribution to the covariance matrix of UE position error through linear transformation.

According to equation (35), the MSE of UE position estimation in the presence of UAV control error can be calculated as follows:
\begin{equation}
MSE_{m,t + 1}^{\left( {{{\bm{\xi }}_{m,t}}} \right)} = {\mathop{\rm tr}\nolimits} \left( {{\bf{Q}}_{\Delta {\bf{p}},m,t + 1}^{\left( {{{\bm{\xi }}_{m,t}}} \right)}} \right).
\end{equation}

\subsection{QoS-Constrained Scheduling Problem}
According to the analysis results described in the previous two subsections, the quality of positioning services in the proposed system is affected mainly by the UAV control error in the next time slot ($t + 1$), while the latter depends on UAV$_{A}$'s operation modes in the current time slot ($t$). Obviously, the schedule of UAV$_{A}$ state sensing is one of the determinant factors of operation modes. In this article, we use a binary vector ${{\bm{\varphi }}_t} = \left[ { \cdots ,{\varphi _{j,t}}, \cdots } \right] \in {\mathbb{Z}^{1 \times {N_A}}}$ to indicate the sensing schedule. ${\varphi _{j,t}} = 1$ means that the $j$-th UAV$_{A}$ is required to perform state sensing in time slot $t$, while ${\varphi _{j,t}} = 0$ means the opposite. If a UAV$_{A}$ is not scheduled in the current time slot, it will of course be controlled in OL mode. However, it is noteworthy that ${\varphi _{j,t}} = 1$ is just a necessary condition for CL mode. The implementation of CL mode also requires the successful transmission of sensing data.

As mentioned at the beginning of Section II, the A2G channel between the $i$-th UAV$_{B}$ and the corresponding terrestrial BS is modeled as a Rayleigh fading channel, whose channel power gain (${\left| {{h_{i,t}}} \right|^2}$) in time slot $t$ can be obtained by CC through channel estimation. We use variable $\theta _{j,t}^i$ ($i \in {\cal N}_B^{{A_j}}$) to represent the blocklength allocated for the transmission of sensing data corresponding to the $j$-th UAV$_{A}$ and $i$-th UAV$_{B}$; Vector ${{\bm{\theta }}_{j,t}} = \left[ { \cdots ,\theta _{j,t}^i, \cdots } \right] \in {\mathbb{R}^{1 \times 3}}$ denotes the blocklength for the transmission of the $j$-th UAV$_{A}$'s sensing data. Then, the success rate of the data transmission corresponding to the $j$-th UAV$_{A}$ and $i$-th UAV$_{B}$ can be calculated with equations introduced in Appendix C and denoted as $P_{j,t}^{i,\left( s \right)}\left( {\theta _{j,t}^i} \right)$. In the proposed system, each UAV$_{A}$ is assigned three UAV$_{B}$ for state sensing, which is the minimum number of anchor nodes required for location estimation. Thus, the probabilities of the two operation modes can be, respectively, expressed as
\begin{equation}
\begin{split}
P_{j,t}^{\left( c \right)}\left( {{\varphi _{j,t}},{{\bm{\theta }}_{j,t}}} \right) &= {\varphi _{j,t}} \cdot \prod\limits_{i \in {\cal N}_B^{{A_j}}} {P_{j,t}^{i,\left( s \right)}\!\left( {\theta _{j,t}^i} \right)} ,\quad\;{\rm closed\raisebox{0mm}{-}loop},\\
P_{j,t}^{\left( o \right)}\left( {{\varphi _{j,t}},{{\bm{\theta }}_{j,t}}} \right) &= 1 - P_{j,t}^{\left( c \right)}\left( {{\varphi _{j,t}},{{\bf{\theta }}_{j,t}}} \right),\quad\quad\;\,{\rm open\raisebox{0mm}{-}loop}.
\end{split}
\end{equation}

Since each UAV$_{A}$ has two operation modes (CL or OL), the vector ${{\bm{\xi }}_{m,t}}$ introduced in subsection B has $G = {2^{N_A^U}}$ possible values, indicating different ``sensing events''. The $G$ possible events are represented by the set ${\cal G} = \left\{ {1, \cdots ,G} \right\}$, and the value of ${{\bm{\xi }}_{m,t}}$ in the $g$-th event ($g \in {\cal G}$) is denoted as ${\bm{\xi }}_{m,t}^g$. Sets ${\cal C}_{m,t}^g$ and ${\cal O}_{m,t}^g$ denote the UAV$_{A}$ operating in CL and OL modes in the $g$-th event, respectively. Then, the expectation of the MSE of the $m$-th UE's position estimate in time slot $t$ can be written as
\begin{equation}
MS{E_{m,t + 1}}\left( {{{\bm{\varphi }}_t},{{\bm{\theta }}_t}} \right) = \sum\limits_{g \in {\cal G}} {\left(\! {MSE_{m,t + 1}^{\left( {{\bm{\xi }}_{m,t}^g} \right)} \cdot P_{m,t}^{\left( {{\bm{\xi }}_{m,t}^g} \right)}\left( {{{\bm{\varphi }}_t},{{\bm{\theta }}_t}} \right)} \right)} ,
\end{equation}
where
\begin{equation}
P_{m,t}^{\left( {{\bm{\xi }}_{m,t}^g} \right)}\left( {{{\bm{\varphi }}_t},{{\bm{\theta }}_t}} \right) = \prod\limits_{j \in {\cal C}_{m,t}^g} {P_{j,t}^{\left( c \right)}\left( {{\varphi _{j,t}},{{\bm{\theta }}_{j,t}}} \right)} \prod\limits_{j \in {\cal O}_{m,t}^g} {P_{j,t}^{\left( o \right)}\left( {{\varphi _{j,t}},{{\bm{\theta }}_{j,t}}} \right)}
\end{equation}
is the probability of the $g$-th sensing event. In the following, $MS{E_{m,t + 1}}\left( {{{\bm{\varphi }}_t},{{\bm{\theta }}_t}} \right)$ calculated by the above equations will be used as the performance metric for positioning services.

In this article, we hope to reduce the resource consumption of UAV$_{A}$ state sensing and control by jointly optimizing the sensing scheduling and blocklength allocation, while ensuring that the quality of positioning services meets UEs' requirements. This goal could be approached by minimizing the total number of symbols used in each time slot, which can be formulated as the following QoS-constrained optimization problem:
\begin{align*}
&\left({{\rm{P1}}} \right):\quad\mathop {\min}\limits_{{{\bm{\varphi }}_t},{{\bm{\theta }}_t}}\,\;\; \sum\limits_{j \in {{\cal N}_A}} {\sum\limits_{i \in {\cal N}_B^{{A_j}}} {\theta _{j,t}^i} } \tag{40}&\\
&\qquad\qquad\;\mbox{s.t.}\quad MS{E_{m,t + 1}}\left( {{\bm{\varphi} _{t}},{{\bm{\theta }}_{j,t}}} \right) \le MSE_m^{Req},\quad\;\; \forall m \in {\cal M}, \tag{41}
\end{align*}
where ${{\bm{\theta }}_t} = \left[ { \cdots \!,{{\bm{\theta }}_{j,t}},\! \cdots } \right] \in {\mathbb{R}^{1 \times \left( {3{N_A}} \right)}}$ ($j \!\in\! {{\cal N}_A}$); Term $\sum\limits_{j \in {{\cal N}_A}} {\sum\limits_{i \in {\cal N}_B^{{A_j}}} {\theta _{j,t}^i} }$ denotes the overall blocklength in time slot $t$; $MSE_m^{Req}$ is the $m$-th UE's requirement on the MSE of positioning services.

\section{Proposed QoS-Qriented Co-Design Solution}
Due to the binary vector ${{\bm{\theta }}_t}$ as well as the nonlinear constraint (41), problem (P1) formulated in the previous section is a mixed-integer nonlinear optimization problem, which is quite difficult to solve optimally, especially when the number of UAV$_{A}$ is large. To make matters worse, in the proposed system, each UE is served by multiple UAV$_{A}$, and each UAV$_{A}$ could also serve multiple UEs. Thus, simply changing the operation mode of one UAV$_{A}$ may not be sufficient to meet the requirement of a certain UE, and may even affect the services for other UEs. The coupling among different UAV$_{A}$ and UEs increases the difficulty of the problem. To solve this problem efficiently, we develop a novel scheme in this section, which divides the original problem into two subproblems, namely the sensing scheduling and blocklength allocation, and solves them successively. The most notable advantage of the proposed scheme is that it decouples the control of each UAV$_{A}$, making it suitable for large-scale UAV systems.

First, by substituting equations (35) and (36) into (38), the expression of $MS{E_{m,t + 1}}\left( {{{\bm{\varphi }}_t},{{\bm{\theta }}_t}} \right)$ can be rewritten as
\begin{equation}\tag{42}
MS{E_{m,t + 1}}\left( {{{\bm{\varphi }}_t},{{\bm{\theta }}_t}} \right) = \sigma _r^2{\mathop{\rm tr}\nolimits} \left( {{\bf{P}}\left( {{{\bf{p}}_m}} \right)} \right) + {\mathop{\rm tr}\nolimits} \left( {{\bf{S}}\left( {{{\bf{p}}_m}} \right){{{\bf{\bar D}}}_{\Delta {\bf{q}},m,t + 1}}{\bf{S}}{{\left( {{{\bf{p}}_m}} \right)}^T}} \right),
\end{equation}
where
\begin{equation}\tag{43}
{{\bf{\bar D}}_{\Delta {\bf{q}},m,t + 1}}\left( {{{\bm{\varphi }}_t},{{\bm{\theta }}_t}} \right) = \sum\limits_{g \in {\cal G}} {\left( {{\bf{D}}_{\Delta {\bf{q}},m,t + 1}^{\left( {{\bm{\xi }}_{m,t}^g} \right)} \cdot P_{m,t}^{\left( {{\bm{\xi }}_{m,t}^g} \right)}\left( {{{\bm{\varphi }}_t},{{\bm{\theta }}_t}} \right)} \right)} ,
\end{equation}
\begin{equation}\tag{44}
{\bf{D}}_{\Delta {\bf{q}},m,t + 1}^{\left( {{\bm{\xi }}_m^g} \right)} = {\mathop{\rm diag}\nolimits} \left( { \ldots ,{\bf{h}}_m^j{{\left( {{{\bf{p}}_m}} \right)}^T}{\bf{Q}}_{\Delta {\bf{q}},j,t + 1}^{\left( {\xi _{m,j,t}^g} \right)}{\bf{h}}_m^j\left( {{{\bf{p}}_m}} \right), \ldots } \right).
\end{equation}
Then, the QoS constraint (41) can be rewritten as
\begin{equation}\tag{45}
{\mathop{\rm tr}\nolimits} \left( {{\bf{S}}\left( {{{\bf{p}}_m}} \right){{{\bf{\bar D}}}_{\Delta {\bf{q}},m,t + 1}}\left( {{{\bm{\varphi }}_t},{{\bm{\theta }}_t}} \right){\bf{S}}{{\left( {{{\bf{p}}_m}} \right)}^T}} \right) \le Th{r_m},
\end{equation}
where $Th{r_m} = MSE_m^{Req} - \sigma _r^2{\mathop{\rm tr}\nolimits} \left( {{\bf{P}}\left( {{{\bf{p}}_m}} \right)} \right)$.

Please note that matrix ${{\bf{\bar D}}_{\Delta {\bf{q}},m,t + 1}}\left( {{{\bm{\varphi }}_t},{{\bm{\theta }}_t}} \right)$ is a diagonal matrix whose diagonal elements denote the expectation of the squared projected length of each UAV$_{A}$'s covariance ellipse in the UAV$_{A}$-UE direction. As can be seen from equation (45), the QoS constraint in problem (P1) only restricts the value obtained through the linear combination of squared projected lengths corresponding to $N_A^U$ UAV$_{A}$ serving each UE. The coupling among UAV$_{A}$ makes it difficult to decide which UAV$_{A}$ should be controlled in CL mode in the current time slot. Thus, we have the following proposition to decouple the control of UAV$_{A}$, so as to reduce the difficulty of sensing scheduling.

\emph{Proposition 1:} Inequality (45) holds for the $m$-th UE if all of the UAV$_{A}$ ($\forall j \in {\cal N}_A^{{U_m}}$) serving this UE fulfill the following condition:
\begin{equation}\tag{46}
{\bf{h}}_m^j {\left( {{{\bf{p}}_m}} \right)^T} \sum\limits_{g \in {\cal G}} {\left( {{\bf{Q}}_{\Delta {\bf{q}},j,t + 1}^{\left( {\xi _{m,j,t}^g} \right)} \cdot P_{m,t}^{\left( {{\bm{\xi }}_m^g} \right)} \left( {{{\bm{\varphi }}_t},{{\bm{\theta }}_t}} \right)} \right)} {\bf{h}}_m^j\left( {{{\bf{p}}_m}} \right) \le \sigma _{\Delta {\bf{q}},m}^2,
\end{equation}
where
\begin{equation}\tag{47}
\sigma _{\Delta {\bf{q}},m}^2 = {{Th{r_m}} \mathord{\left/
 {\vphantom {{Th{r_m}} {{\mathop{\rm tr}\nolimits} \left( {{\bf{P}}\left( {{{\bf{p}}_m}} \right)} \right)}}} \right.
 \kern-\nulldelimiterspace} {{\mathop{\rm tr}\nolimits} \left( {{\bf{P}}\left( {{{\bf{p}}_m}} \right)} \right)}}.
\end{equation}
\begin{proof}
Please refer to Appendix D.
\end{proof}

Please note that condition (46) is sufficient but not necessary for inequality (45) to hold. With Proposition 1, the constraint on the QoS of positioning services is converted into the requirement on the squared projected length of each UAV$_{A}$'s covariance ellipse. Then, we could consider each UAV$_{A}$ individually, and the original problem (P1) can be divided into the following subproblems:
\begin{align*}
&\!\left({{\rm{P2}}} \right)\!:\!\mathop {\min}\limits_{{\varphi _{j,t}},{{\bm{\theta }}_{j,t}}} \sum\limits_{i \in {\cal N}_B^{{A_j}}} {\theta _{j,t}^i} \tag{48}&\\
&\;\,\mbox{s.t.}\;\; {\bf{h}}_m^j{\left(\! {{{\bf{p}}_m}} \!\right)^T}\!\left(\! {P_{j,t}^{\left(\! c \!\right)}\!\left(\! {{\varphi _{j,t}},{{\bm{\theta }}_{j,t}}} \!\right)\!{\bf{Q}}_{\Delta {\bf{q}},j,t \!+\! 1}^{\left( c \right)} + P_{j,t}^{\left(\! o \!\right)}\!\left(\! {{\varphi _{j,t}},{{\bm{\theta }}_{j,t}}} \!\right)\!{\bf{Q}}_{\Delta {\bf{q}},j,t \!+\! 1}^{\left(\! o \!\right)}} \!\right)\!{\bf{h}}_m^j\!\left(\! {{{\bf{p}}_m}} \!\right) \!\le\! \sigma _{\Delta {\bf{q}},m}^2, \;\forall m \!\in\! {{\cal M}_j}, \tag{49}
\end{align*}
where ${{\cal M}_j}$ is the set of UEs served by the $j$-th UAV$_{A}$.

Although the number of variables in a single problem has been greatly reduced, problem (P2) is still a complex mixed-integer nonlinear problem. Thus, later in this section, we use two strategies to solve the subproblems of sensing scheduling and blocklength allocation in problem (P2) successively.

The main idea of our strategy for sensing scheduling can be summarized as follows: If the stability of a UAV$_{A}$ in OL mode meets the requirement, the condition (46) could be satisfied. Therefore, we will try to control a UAV$_{A}$ in CL mode if its predicted stability in OL mode approaches the limit. We use the following variable to indicate the gap between the squared projected lengths of the $j$-th UAV$_{A}$ in OL mode and UEs' requirements:
\begin{equation}\tag{50}
{\kappa _{j,t}} = \mathop {\max }\limits_{m \in {{\cal M}_j}} \left\{ {{\bf{h}}_m^j{{\left( {{{\bf{p}}_m}} \right)}^T}{\bf{Q}}_{\Delta {\bf{q}},j,t + 1}^{\left( o \right)}{\bf{h}}_m^j\left( {{{\bf{p}}_m}} \right) - {\lambda ^2} \cdot \sigma _{\Delta {\bf{q}},m}^2} \right\},
\end{equation}
where $\lambda $ is a scaling factor representing our tolerance for the increase in UAV position uncertainty caused by OL modes. Then, our strategy for sensing scheduling can be expressed as
\begin{equation}\tag{51}
{\varphi _{j,t}} = \left\{ {\begin{array}{*{20}{c}}
\begin{aligned}
&\!{1},&&\qquad{\rm if}\;\,{\kappa _{j,t}} \ge 0,\\
&\!{0},&&\qquad{\rm otherwise}.
\end{aligned}
\end{array}} \right.
\end{equation}

If the $j$-th UAV$_{A}$ is chosen to be controlled in CL mode, we hope that the corresponding state sensing could be carried out with a relative high success rate, so as to avoid the UAV position uncertainty exceeding the tolerable limit due to frequent transmission failures. Since this article is just a preliminary attempt to introduce the concept of SCC co-design into UAV-enabled positioning, we adopt a simple strategy to allocate the blocklength for data transmission. In our strategy, the success rate of state sensing is set to a fixed value $P_{Req}^{\left( c \right)}$, and the transmission of sensing data obtained by different UAV$_{B}$ has the same success rate. Then, the blocklength allocated to the $i$-th UAV$_{B}$ ($i \in {\cal N}_B^{{A_j}}$) for transmitting the $j$-th UAV$_{A}$'s sensing data can be calculated as
\begin{equation}\tag{52}
\theta _{j,t}^i = \theta\! \left( {{{\left( {P_{Req}^{\left( c \right)}} \right)}^{{1 \mathord{\left/
 {\vphantom {1 3}} \right.
 \kern-\nulldelimiterspace} 3}}},{\gamma _{i,t}}} \right) \cdot {\varphi _{j,t}},
\end{equation}
where $\theta \left(  \cdot  \right)$ is the blocklength calculation function, and ${\gamma _{i,t}}$ denotes the signal-to-noise ratio (SNR) of the $i$-th UAV$_{B}$'s signal at the corresponding BS. The expressions of $\theta \left(  \cdot  \right)$ and ${\gamma _{i,t}}$ are derived in Appendix C.

With the scheme and strategies introduced in this section, the proposed system is supposed to work efficiently and provide positioning services that meet UEs' requirements.

\section{Numerical Results}
In this section, we conduct a series of simulation experiments and provide the corresponding numerical results to validate the performance of the proposed system. First, the importance of SCC co-design for UAV-enabled positioning and the validity of our system are verified by an experiment. Then, we compare the proposed co-design scheme with two benchmark schemes, namely the ``continuous scheme'' and ``periodic scheme'', to reflect the superiority of our scheme. Finally, some key factors affecting the performance of the proposed system and their influence on positioning services are quantitatively analyzed to provide guidance for the application of our system in practical situations.
\begin{figure}[!t]
\centering
\includegraphics[height=1.65in,width=3.45in]{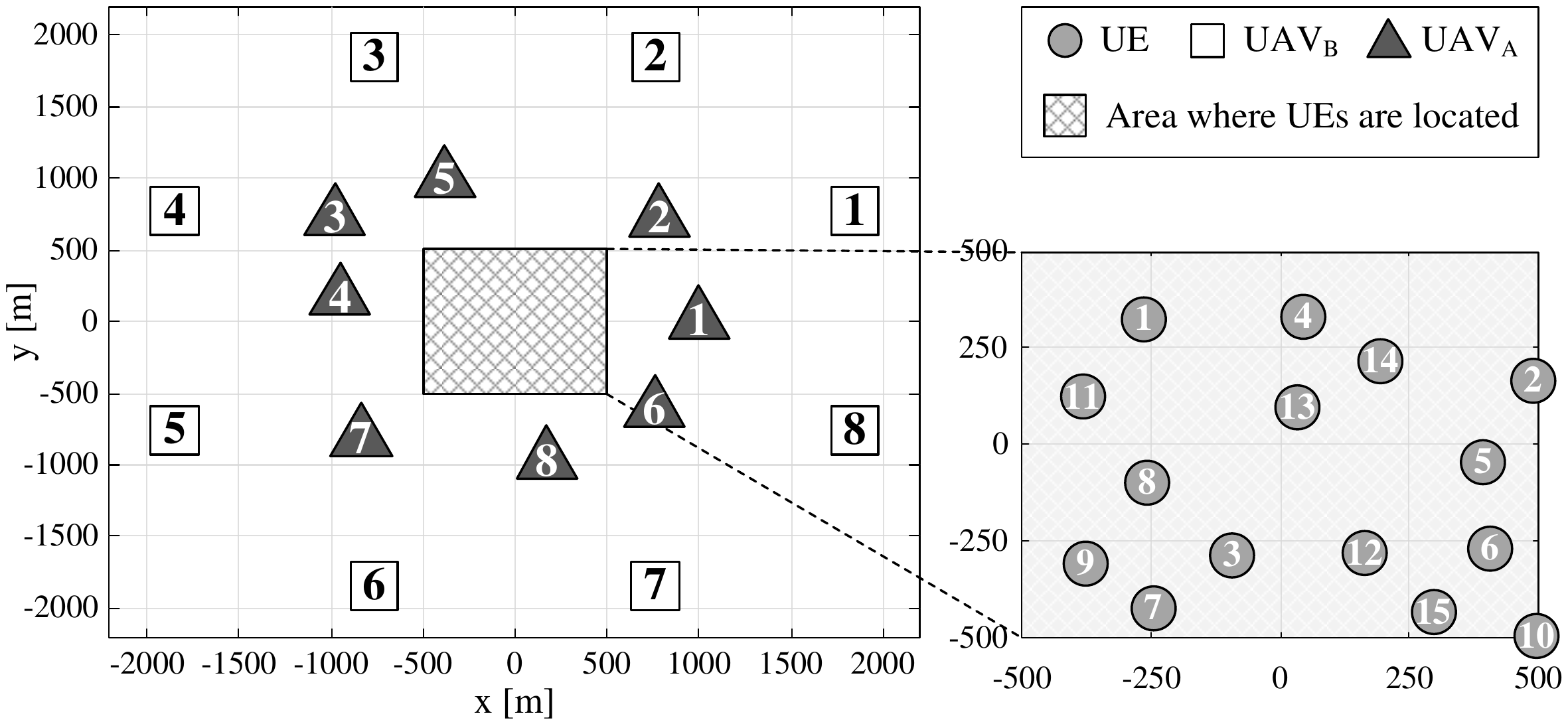}
\caption{Test scenario for numerical evaluation.}
\label{fig_4}
\end{figure}

Fig. 4 shows the test scenario used for performance evaluation and comparison in this section, which consists of 8 UAV$_{A}$ (${N_A} = 8$), 8 UAV$_{B}$ (${N_B} = 8$) and 15 UEs ($M = 15$). The origin of Cartesian coordinates is set at the center of the test scenario. UAV$_{B}$ are uniformly deployed on the circle of radius 2 km centered at the origin, and the distance between each UAV$_{B}$ and its corresponding BS is 1 km. The horizontal coordinates of the 8 UAV$_{A}$'s HPs are set to ${\bf{q}}_1^ \circ  \!=\! {\left[ {1000,0} \right]^T}$, ${\bf{q}}_2^ \circ  \!=\! {\left[ {785,715} \right]^T}$, ${\bf{q}}_3^ \circ  \!=\! {\left[ { - 980,724} \right]^T}$, ${\bf{q}}_4^ \circ  \!=\! {\left[ { - 951,164} \right]^T}$, ${\bf{q}}_5^ \circ  \!=\! {\left[ { - 382,990} \right]^T}$, ${\bf{q}}_6^ \circ  \!=\! {\left[ {758, - 624} \right]^T}$, ${\bf{q}}_7^ \circ  \!=\! {\left[ { - 836, - 820} \right]^T}$ and ${\bf{q}}_8^ \circ  \!=\! {\left[ {172, - 977} \right]^T}$. UEs are randomly located within a square with center at the coordinate origin and side length of 1 km. When assigning $N_A^U$ UAV$_{A}$ to each UE, we select the subset of UAV$_{A}$ with the minimum horizontal dilution of precision (HDOP) \cite{UAV_Avai}. The same criterion is also used for the selection of the three UAV$_{B}$ responsible for sensing each UAV$_{A}$'s state.

The key simulation parameters used in this section are summarized as follows: All UAVs in the proposed system, including UAV$_{A}$ and UAV$_{B}$, hover at the same altitude ${h_v} \!=\! 50$ m and have the same transmit power ${P_t} \!=\! 30$ dBm; The noise power at terrestrial BSs is ${N_0} \!=\!  - 107$ dBm; The number of UAV$_{A}$ serving each UE is $N_A^U = 3$; The variances of inter-UAV and UAV$_{A}$-UE distance measurements are $\sigma _d^2 = 1$ ${{\rm m^2}}$ and $\sigma _r^2 = 1$ ${{\rm m^2}}$, respectively; UEs have the same requirement on the MSE of positioning services, i.e., $MS{E_{Req,m}} = {10^2}$ ${{\rm m^2}}$ ($\forall m \in {\cal M}$). The covariance matrices of velocity and acceleration measurements are set to ${{\bf{R}}_{\bf{v}}} = \left( {{{0.5}^2}} \right) \cdot {{\bf{I}}_2}$ and ${{\bf{R}}_{\bf{a}}} = \left( {{{0.1}^2}} \right) \cdot {{\bf{I}}_2}$, respectively; The length of each time slot is $\Delta t = 1$ s; The time constant of lag of each UAV$_{A}$ in responding commanded acceleration is $\rho  = 10$ ms; The continuous-time acceleration process noise intensity in $x$- and $y$-directions are $\varsigma _x^2 = {0.5^2}$ ${{{m^2}} \mathord{\left/
 {\vphantom {{{m^2}} {{s^3}}}} \right.
 \kern-\nulldelimiterspace} {{s^3}}}$ and $\varsigma _y^2 = {0.5^2}$ ${{{m^2}} \mathord{\left/
 {\vphantom {{{m^2}} {{s^3}}}} \right.
 \kern-\nulldelimiterspace} {{s^3}}}$, respectively; The feedback gain matrix ${{\bf{K}}_j}$ for UAV control is generated based on the standard linear quadratic regulator (LQR) control; The scaling factor and success rate in the proposed co-design scheme are set to $\lambda  = 0.8$ and $P_{Req}^{\left( c \right)} = 0.95$, respectively.

\subsection{The Importance of Co-Design in UAV-Enabled Positioning}
In this subsection, we successively apply two UAV-enabled positioning systems, namely the conventional system without considering state sensing and control as well as the proposed co-design system, in the test scenario shown in Fig. 4. The proposed system operates according to the scheme described in Section IV, while the dynamics model of the conventional system is obtained by removing the term ${{\bf{B}}_j}{{\bf{u}}_{j,t}}$ from equation (11). The running time of both systems is 60 time slots (1 min), and the corresponding simulation results are shown in Fig. 5 and Fig. 6. For the simplicity of data presentation, this subsection only presents the simulation results corresponding to a part of UAV$_{A}$ and UEs, which will not reduce the generality of the analysis results.
\begin{figure}[!t]
\centering
\includegraphics[height=1.9in,width=3.45in]{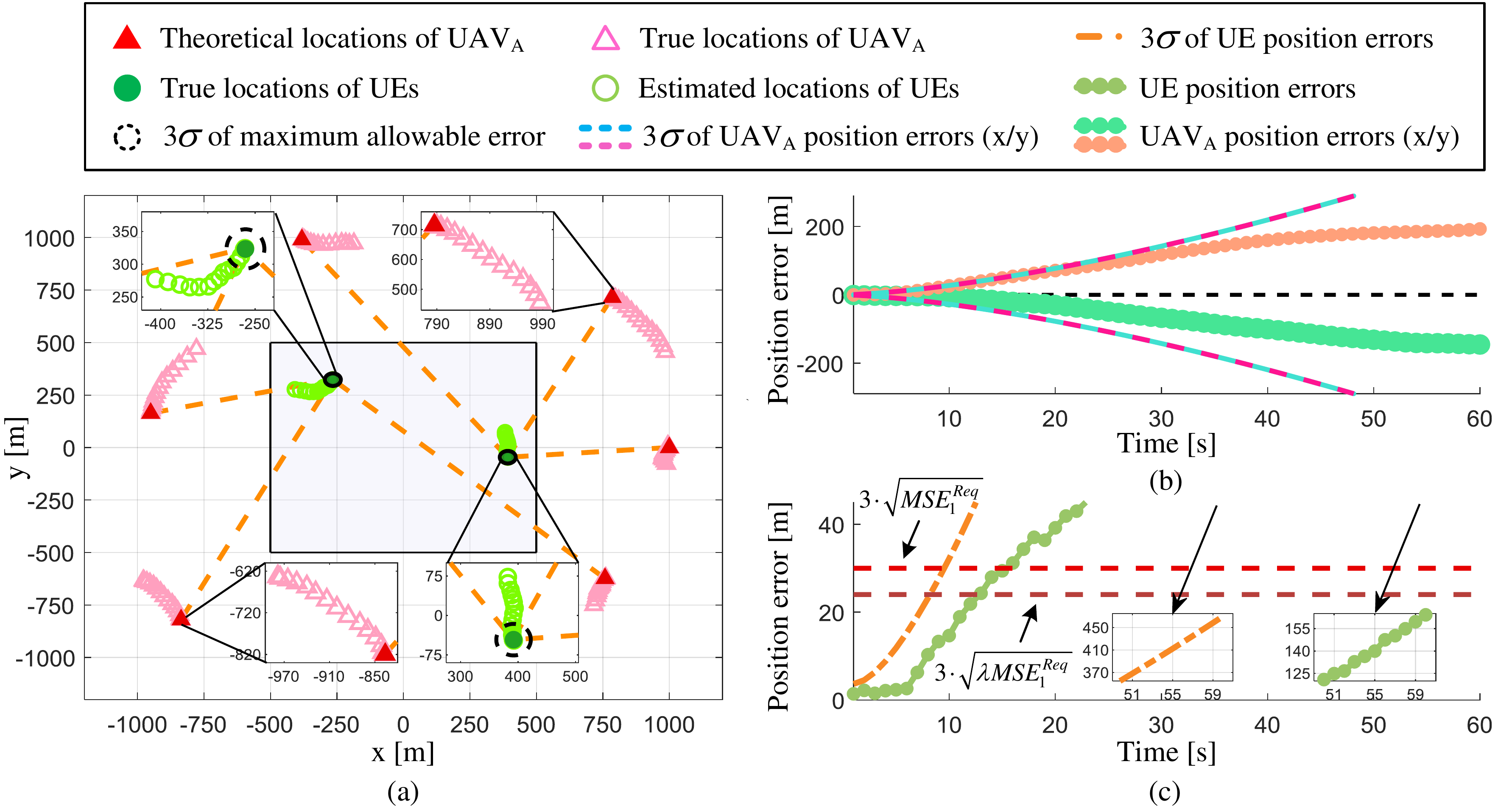}
\caption{Test results of the system without UAV state sensing: Variations of (a) system status, (b) UAV control error (${\rm UAV_{A}}$ 7) and (c) UE positioning performance (UE 1).}
\label{fig_5}
\end{figure}

Fig. 5(a) and Fig. 6(a) show the variations of the two systems' statuses over time, including the variations of UAV$_{A}$'s true locations and UEs' location estimates. The location estimate of each UE is obtained through the ILS method introduced in Section II.C, and the estimation is performed once in each time slot. It can be seen that UAV$_{A}$ in the conventional system deviate significantly from their HPs. According to Fig. 5(b), these deviations reach hundreds of meters in both $x$- and $y$-directions after 40 s, almost threatening the system safety. In terms of QoS, as can be seen from Fig. 5(c), the theoretical MSE of the conventional system's positioning services exceeds the ${10^2}$ ${m^2}$ required by UEs only 10 s after the experiment starts. About 5 s later, the position estimation error of ILS also exceeds the tolerable limit. At the end of the experiment, the theoretical MSE and position error of the conventional system have reached the astonishing values of ${156.7^2}$ ${m^2}$ and $164.2$ m, respectively. These phenomena indicate that sensing and control functionalities are indispensable for the implementation of UAV-enabled positioning.
\begin{figure}[!t]
\centering
\includegraphics[height=1.9in,width=3.45in]{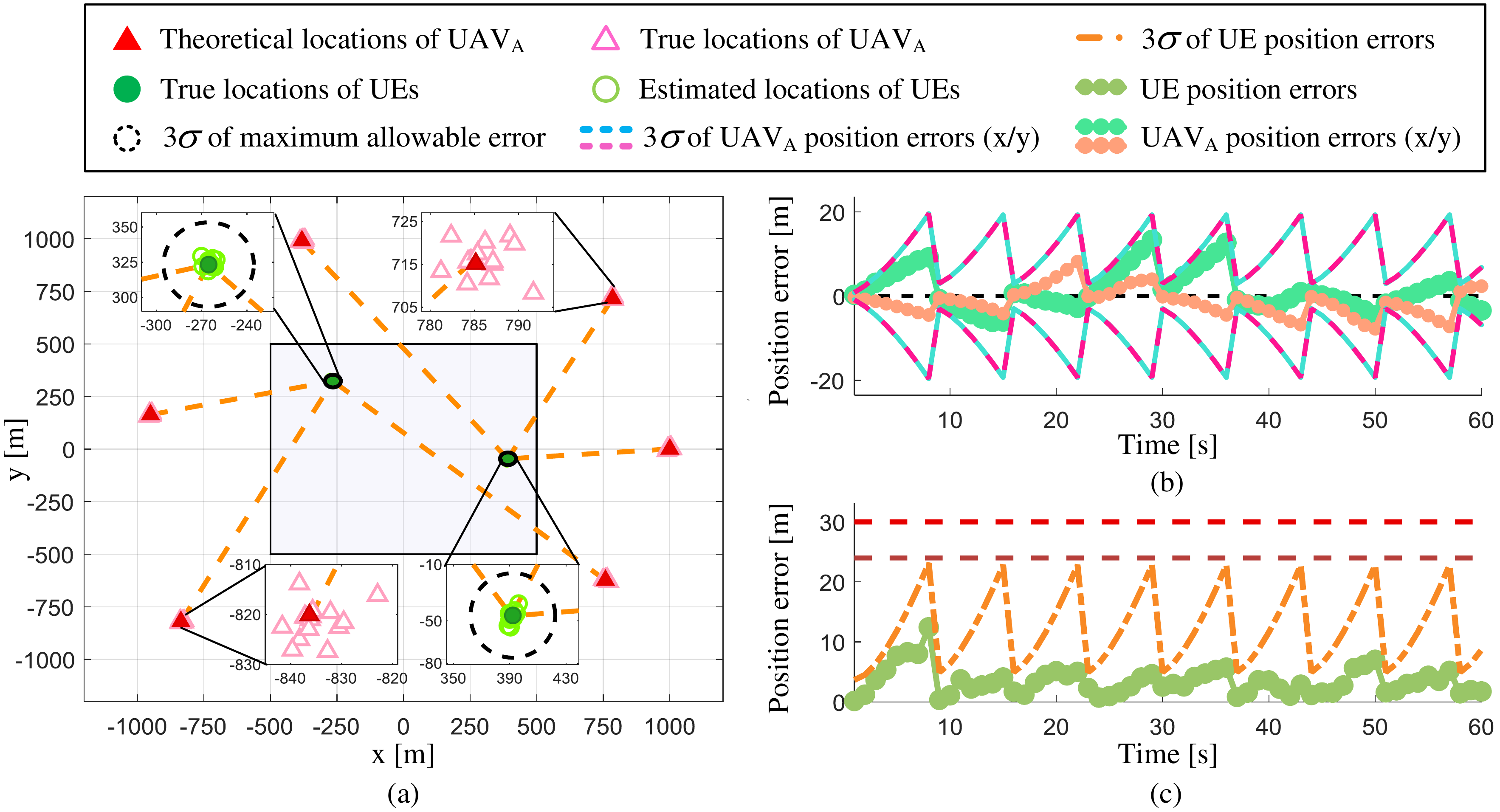}
\caption{Test results of the proposed system using the co-design scheme: Variations of (a) system status, (b) UAV control error (${\rm UAV_{A}}$ 7) and (c) UE positioning performance (UE 1).}
\label{fig_6}
\end{figure}

As shown in Fig. 6(a), during the experiment, UAV$_{A}$ in the proposed system always stay in the vicinity of their HPs, and the position estimation error of each UE is also kept within an acceptable range. According to Fig. 6(b), the proposed system maintains the deviations between UAV$_{A}$'s true location and its HP in both $x$- and $y$-directions within the range of $ - 20.00 \sim 20.00$ m. The variation of UAV control error in Fig. 6(b) can be interpreted as follows: When the position uncertainty of the UAV$_{A}$ is relatively small, such as $1 \!\sim\! 7$ s in this experiment, UAV$_{A}$ is controlled in OL mode to reduce resource consumption, resulting in an increase in UAV control error. If the UAV position uncertainty approaches the limit defined by inequality (46), the proposed system allocates wireless resources for state sensing to the UAV$_{A}$ so that it can be controlled in CL mode. Therefore, the UAV control errors in our system fluctuate dynamically with time. Moreover, as can be seen from Fig. 6(c), the theoretical MSE of our system's positioning services exhibits a variation similar to the variance of the UAV control error. This phenomenon is reasonable because, according to equation (35), the position uncertainty of UAV$_{A}$ is one of the main factors affecting the performance of UE positioning. It is noteworthy that although the quality of positioning services fluctuates with time, the proposed system can always meet UEs' requirements. During the experiment, the maximum theoretical MSE and position estimation error are ${7.717^2}$ ${m^2}$ and $12.44$ m, respectively. The above results demonstrate the feasibility and validity of our system and scheme.

The simulation results in this subsection reflect the importance of state sensing and control for UAV-enabled positioning, as well as the validity of the proposed co-design system. In addition, it can be seen that in both UAV control and UE position estimation, the measured error rarely exceeds three times the theoretical root-mean-square error (RMSE), which demonstrates the correctness of the performance analysis in Section III.

\subsection{The Superiority of Co-Design over the Continuous Scheme}
In this subsection, we test and compare the performance of our scheme introduced in Section IV and the ``continuous scheme'' commonly used in existing systems to illustrate the superiority of SCC co-design. In the continuous scheme, the state sensing for each UAV$_{A}$ is performed in each time slot, i.e., ${\varphi _{j,t}} = 1$ ($\forall j \in {{\cal N}_A}$, $\forall t$). The blocklength in the continuous scheme is determined by the same strategy shown in equation (52). Fig. 7 and Fig. 8 show the numerical results obtained in this experiment.
\begin{figure}[!t]
\centering
\includegraphics[height=2.5in,width=3.45in]{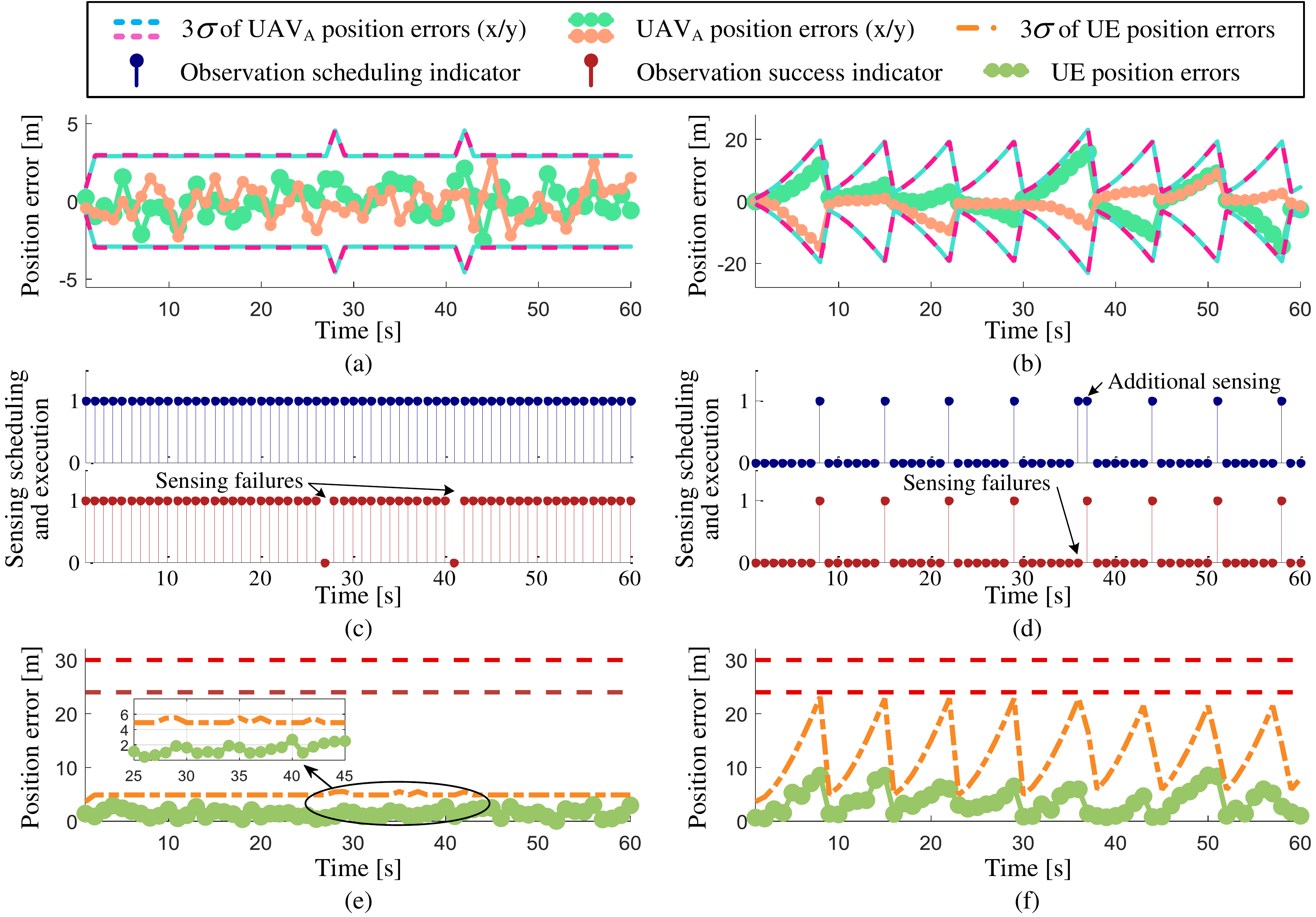}
\caption{Performance comparison between the continuous and proposed schemes: UAV control error (${\rm UAV_{A}}$ 7) of the (a) continuous and (b) proposed schemes; Sensing scheduling and execution status (${\rm UAV_{A}}$ 7) of the (c) continuous and (d) proposed schemes; UE positioning performance (UE 1) of the (e) continuous and (f) proposed schemes}
\label{fig_7}
\end{figure}

As can be seen from Fig. 7(a) and (b), the UAV control error of the continuous scheme is basically maintained within the range of $ - 2.973 \sim 2.973$ m, while that of the proposed system is mostly between $ - 19.33$ m and $19.33$ m. Please note that in the 28 s and 42 s of the operation of the continuous scheme, the envelope of UAV control error expands significantly and reaches $ \pm 4.607$ m. According to the sensing scheduling and execution results shown in Fig. 7(c), the major cause of these phenomena is the unexpected OL control due to the failures of sensing data transmission in 27 s and 41 s. As shown in Fig. 7(b) and (d), a similar phenomenon also occurs in the 37 s of the operation of the proposed scheme. Fortunately, the proposed scheme performs additional state sensing in the 37 s, avoiding the further increase in the variance of UAV control error. This result indicates the proposed scheme's adaptability to transmission failures. Although it seems that the continuous scheme has much smaller UAV control errors than the proposed scheme, the data in Fig. 7(e) and (f) show that such high control accuracy is completely unnecessary for positioning services. According to Fig. 7(e), during the experiment, the maximum theoretical MSE and position estimation error of the continuous scheme's positioning services are ${1.847^2}$ ${m^2}$ and $3.023$ m, which are far below the UEs' tolerable limits. In terms of the proposed system (Fig. 7(f)), its maximum theoretical MSE and position error are ${7.717^2}$ ${m^2}$ and $8.537$ m, which are much larger than those of the continuous scheme but still meet the UEs' requirements.
\begin{figure}[!t]
\centering
\includegraphics[height=1.9in,width=3.45in]{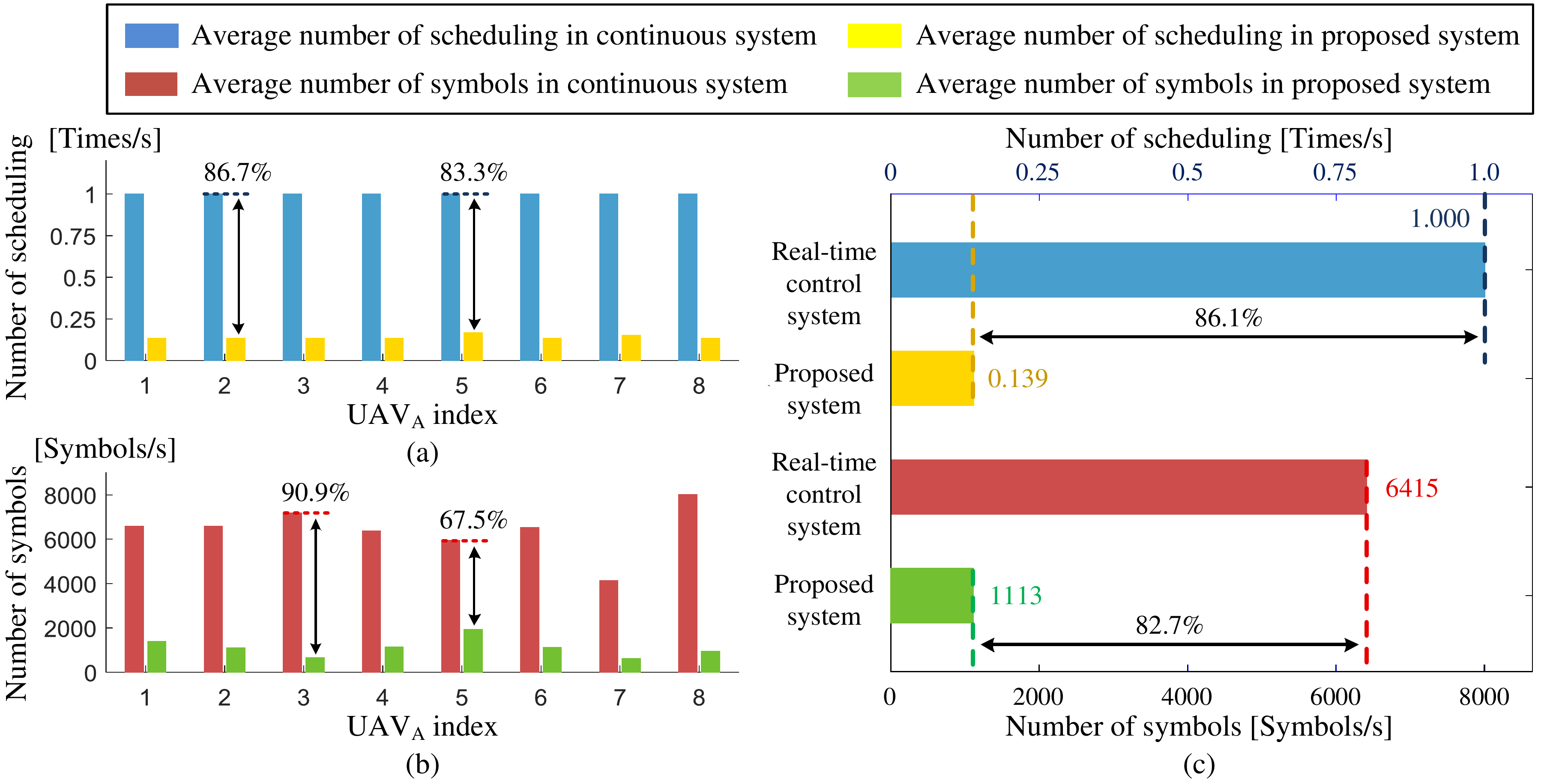}
\caption{Performance comparison between the continuous and proposed schemes: Number of (a) scheduling and (b) symbols corresponding to each ${\rm UAV_{A}}$ in two schemes; (c) average number of scheduling and symbols among all ${\rm UAV_{A}}$ in two schemes.}
\label{fig_8}
\end{figure}

Since both of the schemes tested in this subsection could satisfy UEs' QoS demands, we turn to compare their resource consumption. The average number of scheduling (times/s) and the average number of symbols (symbols/s) used for data transmission are chosen as the metrics for evaluating the resource efficiency of each scheme. Fig. 8 shows the comparison of the resource consumption of the two schemes. As can be seen from Fig. 8(a), compared with the continuous scheme, the proposed scheme significantly reduces the average number of scheduling for each UAV$_{A}$. The reduction in the number of scheduling is between $83.3\% $ and $86.7\% $. Moreover, according to Fig. 8(b), the proposed scheme also greatly reduces the number of symbols required for the state sensing of each UAV$_{A}$, and the reduction is within the range of $67.5\%  \sim 90.9\% $. Fig. 8(c) shows the resource consumption of the two schemes from an overall system perspective. For the overall system, the proposed scheme reduces the number of scheduling and symbols for positioning services by $86.1\% $ and $82.7\% $, respectively.

Based on the above numerical results, the superiority of the proposed SCC co-design scheme over the commonly used continuous scheme can be summarized as follows: The capabilities of state sensing and prediction in the proposed scheme enable the system to schedule and allocate resources flexibly according to the system status, thereby improving the resource efficiency of positioning services while ensuring that QoS meets UEs' requirements.

\subsection{The Superiority of Co-Design over the Periodic Scheme}
The simulation results presented in the previous two subsections (especially Fig. 6(b) and Fig. 7(d)) may give the illusion that the state sensing in the proposed scheme is performed periodically. Therefore, in this subsection, we compare our SCC co-design scheme with the periodic scheme to illustrate the differences between the two schemes as well as the superiority of the proposed scheme. In the periodic scheme, the state sensing for each UAV$_{A}$ is performed periodically at a constant time interval, regardless of whether the sensing data is successfully transmitted. For the parameter setting mentioned at the beginning of this section, this time interval is set to 7 time slots (i.e., 7 s). We set up two test situations for the comparison of the two schemes, namely the low-success-rate situation and the intermittent-link-blocking situation. In each situation, the running time of both schemes is 120 time slots (2 min).
\begin{figure}[!t]
\centering
\includegraphics[height=1.8in,width=3.45in]{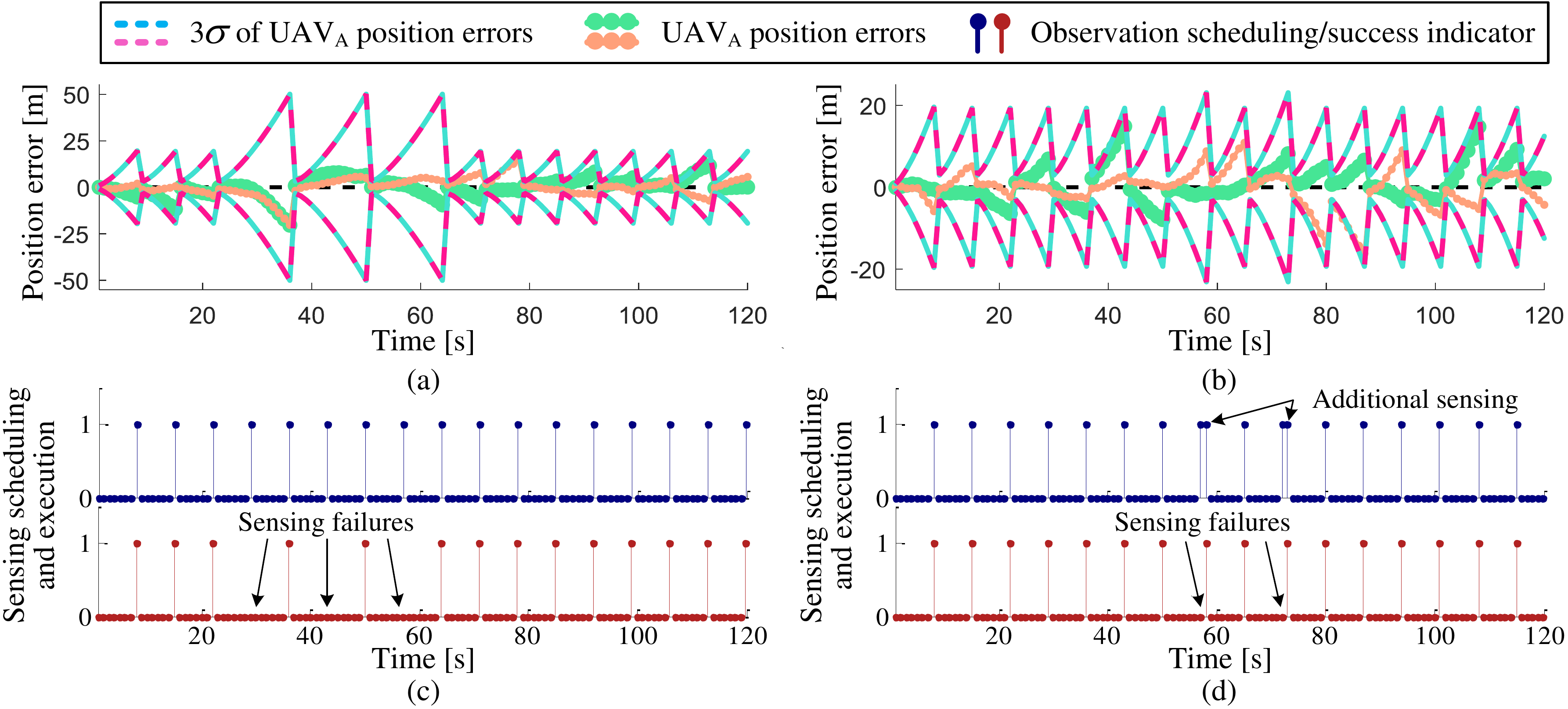}
\caption{Performance comparison between the periodic and proposed schemes under low-success-rate condition: UAV control error (${\rm UAV_{A}}$ 7) of the (a) periodic and (b) proposed schemes; Sensing scheduling and execution status (${\rm UAV_{A}}$ 7) of the (c) periodic and (d) proposed schemes.}
\label{fig_9}
\end{figure}

In the low-success-rate situation, the success rate of state sensing ($P_{Req}^{\left( c \right)}$) is reduced from 0.95 to 0.70, so as to simulate frequent transmission failures due to limited wireless resources and fading channels. The values of other parameters are the same as those mentioned at the beginning of this section. Fig. 9 and Fig. 10 show the performance of the two schemes in this test situation. As can be seen from Fig. 9(c), in the 29 s, 43 s and 57 s of the operation of the periodic scheme, the state sensing for the UAV$_{A}$ is scheduled but not successfully executed. According to Fig. 9(a), the envelope of UAV control error continues to expand after these moments until state sensing is scheduled and executed successfully in the next ``sensing occasion''. During the experiment, the size of the envelope of UAV control error in the periodic scheme reaches a maximum of $ \pm 50.05$ m. For the proposed scheme, the deviation between UAV$_{A}$'s true location and its HP is basically maintained within the range of $ - 19.33 \sim 19.33$ m, as shown in Fig. 9(b). Although sensing failures also occur in the proposed scheme (57 s and 72 s in Fig. 9(d)) and lead to temporary expansions of the envelope of UAV control error, these trends are quickly reversed by performing additional state sensing in the next time slots (58 s and 73 s). The differences in sensing scheduling between the two schemes also affect the quality of their positioning services. As shown in Fig. 10(a), during the experiment, the theoretical MSE of the periodic scheme exceeds the tolerable limit several times, resulting in ``service failures''. On the contrary, as can be seen from Fig. 10(b), the performance of the proposed scheme's positioning services always meets UEs' requirements. According to the sensing scheduling and execution results shown in Fig. 10(c) and (d), the main reason for these phenomena is that the proposed scheme has the ability to perform additional state sensing according to the system status, while the periodic scheme does not. Therefore, in the low-success-rate situation, the service of the proposed SCC co-design scheme is much more reliable than that of the periodic scheme.
\begin{figure}[!t]
\centering
\includegraphics[height=1.7in,width=3.45in]{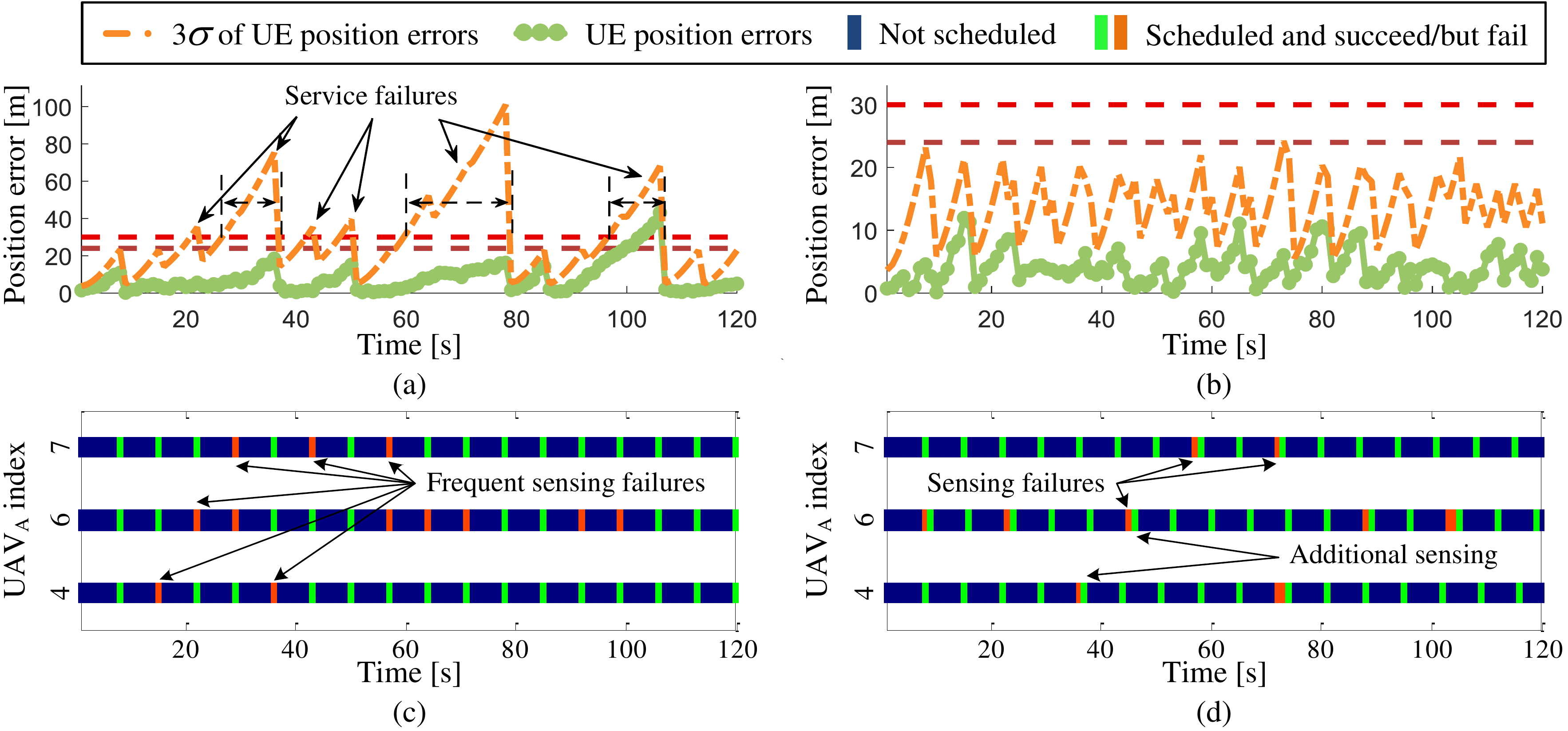}
\caption{Performance comparison between the periodic and proposed schemes under low-success-rate condition: UE positioning performance (UE 1) of the (a) periodic and (b) proposed schemes; Sensing scheduling and execution status (${\rm UAV_{A}}$ 4, 6 and 7 that serve UE 1) of the (c) periodic and (d) proposed schemes.}
\label{fig_10}
\end{figure}
\begin{figure}[!t]
\centering
\includegraphics[height=2.6in,width=3.45in]{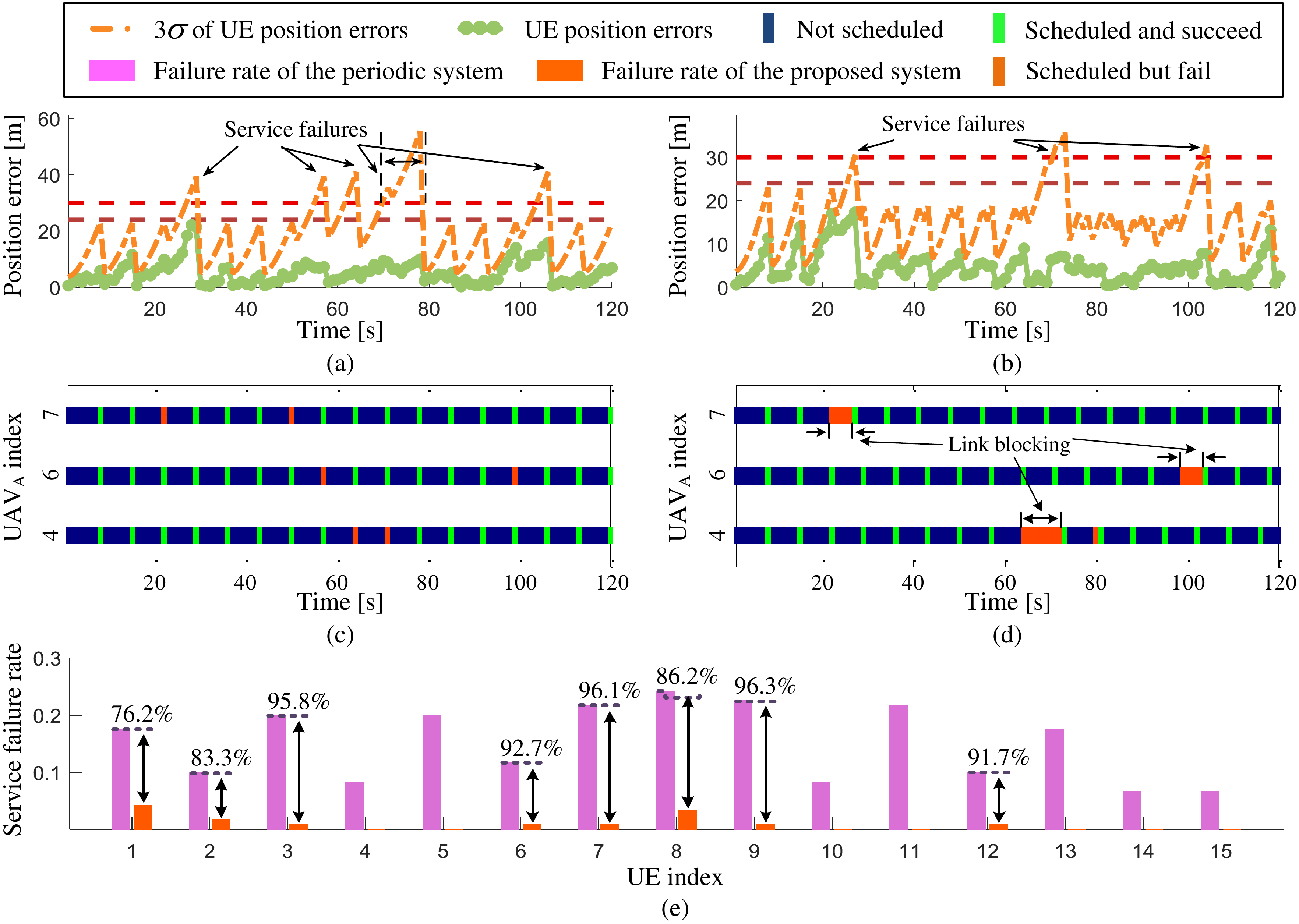}
\caption{Performance comparison between the periodic and proposed schemes under intermittent-link-blocking condition: UE positioning performance (UE 1) of the (a) periodic and (b) proposed schemes; Sensing scheduling and execution status (${\rm UAV_{A}}$ 4, 6 and 7 that serve UE 1) of the (c) periodic and (d) proposed schemes; (e) Service failure rate corresponding to each UE in two schemes}
\label{fig_11}
\end{figure}

The term ``intermittent-link-blocking'' refers to a situation where one or more UAV$_{A}$ are unable to perform state sensing within a short period of time due to equipment failure or other reasons. In this experiment, we assume that the sensing links for UAV$_{A}$ 4, 6 and 7 are blocked in the time periods of $63 \sim 72$ s, $98 \sim 103$ s and $20 \sim 24$ s, respectively. Fig. 11 shows the test results of the two schemes obtained in this situation. As can be seen from Fig. 11 (a) and (b), service failures occur during the operation of both schemes. However, it is worth noting that the number and duration of service failures in the proposed scheme are much smaller than those in the periodic scheme. Moreover, during the experiment, the maximum theoretical MSE of the periodic and proposed schemes is ${18.47^2}$ ${m^2}$ and ${12.01^2}$ ${m^2}$, respectively. Thus, compared with the periodic scheme, the proposed scheme also significantly reduces the impact of link blocking on the quality of positioning services. In Fig. 11(e), the service failure rate of each UE in the experiment is calculated. It can be seen clearly that the proposed scheme greatly reduces the service failure rate of each of the 15 UEs in the system, and the reduction ranges from $76.2\% $ to $96.3\% $. The data in Fig. 11(c) and (d) illustrate that the proposed scheme's ability to perform additional state sensing is also the major cause of these phenomena.

From the above experimental results and analysis, it can be concluded that the capabilities of state sensing and prediction enable our proposed SCC co-design scheme to achieve much better service reliability than the periodic scheme in harsh situations.

\subsection{Key Factors Affecting System Performance}
The validity of the proposed system and its superiority over the benchmark schemes have been demonstrated in the previous subsections. Then, in this subsection, we conduct two experiments to analyze the key factors affecting the performance of the proposed system and their influence on positioning services. Specifically, we consider three factors, namely the tolerable limit of the theoretical MSE of positioning services ($MS{E_{Req,m}}$), the scaling factor ($\lambda $) and success rate ($P_{Req}^{\left( c \right)}$) in the proposed scheme. The system performance is evaluated by the following three metrics: service failure rate, average number of scheduling and average number of symbols. In order to make the experimental results statistically significant, the running time of our system in each experiment is set to $1 \times {10^6}$ time slots (about 11.6 days).
\begin{figure}[!t]
\centering
\includegraphics[height=1.4in,width=3.45in]{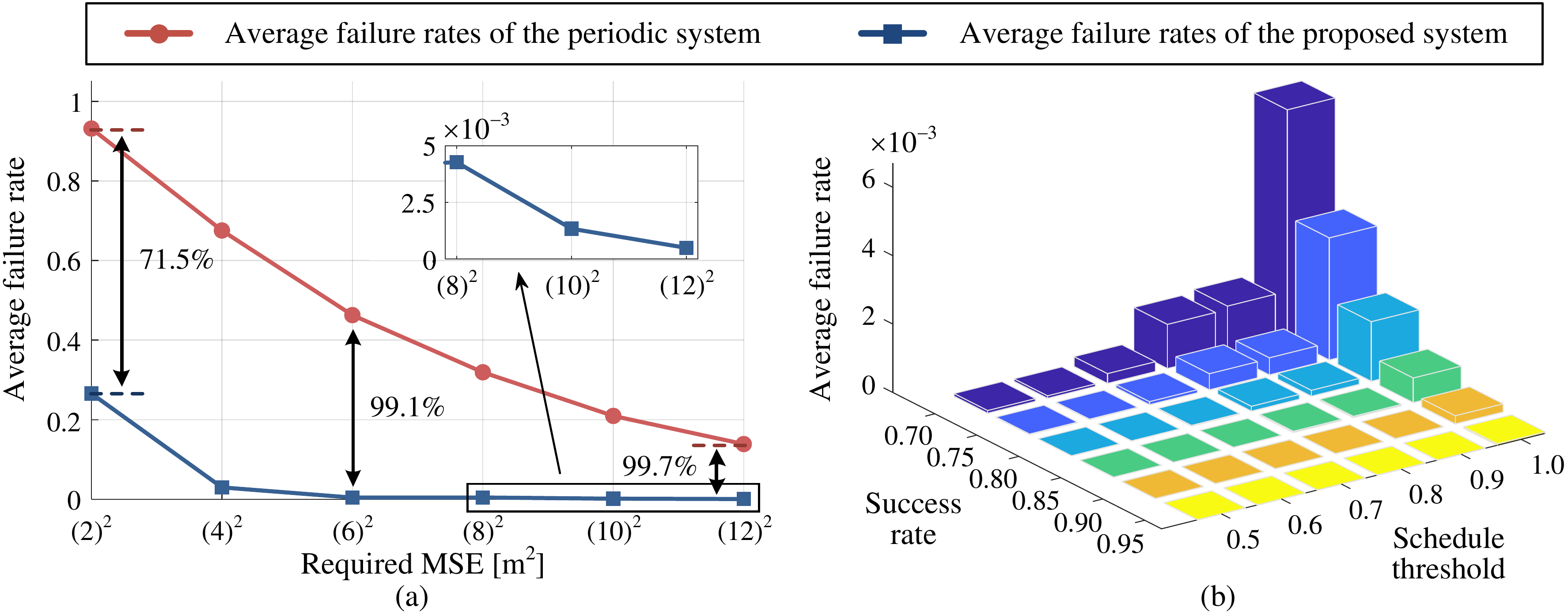}
\caption{Variations of service failure rate with (a) Required RMSE, (b) success rate and schedule threshold.}
\label{fig_12}
\end{figure}
\begin{figure}[!t]
\centering
\includegraphics[height=1.4in,width=3.45in]{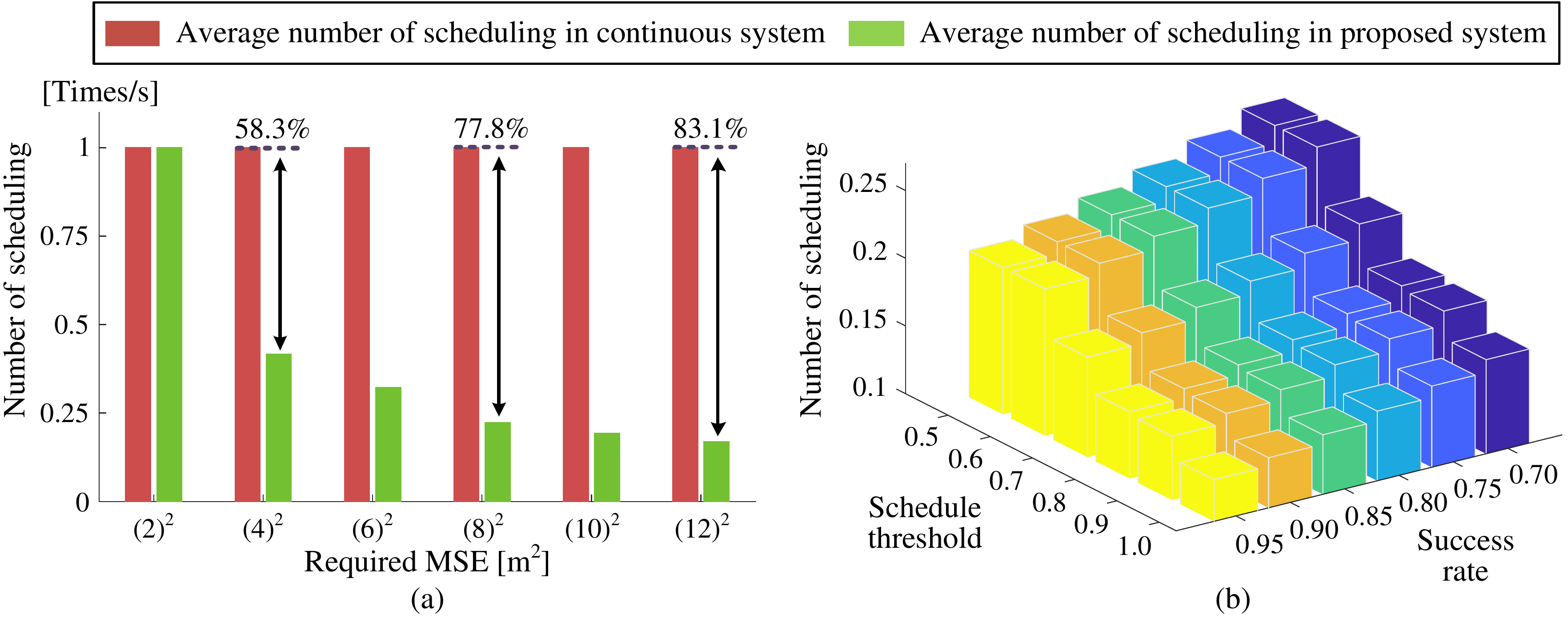}
\caption{Variations of average number of scheduling with (a) Required RMSE, (b) success rate and schedule threshold.}
\label{fig_13}
\end{figure}

In the first experiment, we analyze the influence of UEs' tolerable limit ($MS{E_{Req,m}}$) on system performance. During the simulation, the value of $MS{E_{Req,m}}$ increases from ${2^2}$ ${m^2}$ to ${12^2}$ ${m^2}$, while the values of other parameters are set according to the parameter setting mentioned at the beginning of this section. Fig. 12(a), Fig. 13(a) and Fig. 14(a) show the variations of the three performance metrics with the value of $MS{E_{Req,m}}$. As can be seen from Fig. 12(a), the service failure rate of the proposed system decreases with the increase of $MS{E_{Req,m}}$. Moreover, the advantage of the proposed system over the periodic system in service reliability becomes more significant as $MS{E_{Req,m}}$ increases. When $MS{E_{Req,m}} = {2^2}$ ${m^2}$, the service failure rate of the proposed system is $71.5\% $ lower than that of the periodic system, and this ratio increases to $99.7\% $ after the value of $MS{E_{Req,m}}$ becomes ${12^2}$ ${m^2}$. In terms of the average number of scheduling and symbols, according to Fig. 13(a) and Fig. 14(a), these two performance metrics of the proposed system also decrease with the increase of $MS{E_{Req,m}}$. Since the resource consumption of the continuous system remains basically unchanged during the experiment, the reduction of the resource consumption in the proposed system highlights its advantage in resource efficiency. For example, when $MS{E_{Req,m}} = {2^2}$ ${m^2}$, the proposed system has the same resource consumption as the continuous system. However, when the value of $MS{E_{Req,m}}$ increases to ${12^2}$ ${m^2}$, the average number of scheduling and average number of symbols in the proposed system are $83.1\% $ and $82.6\% $ less than those in the continuous system. The main reason for these phenomena is that the proposed system has the ability to adjust the frequency of state sensing flexibly according to UEs' requirements, while the continuous system does not. The above experimental results indicate that the proposed system has more significant advantages in service reliability and resource efficiency when UEs have relatively loose requirements on position accuracy.
\begin{figure}[!t]
\centering
\includegraphics[height=1.4in,width=3.45in]{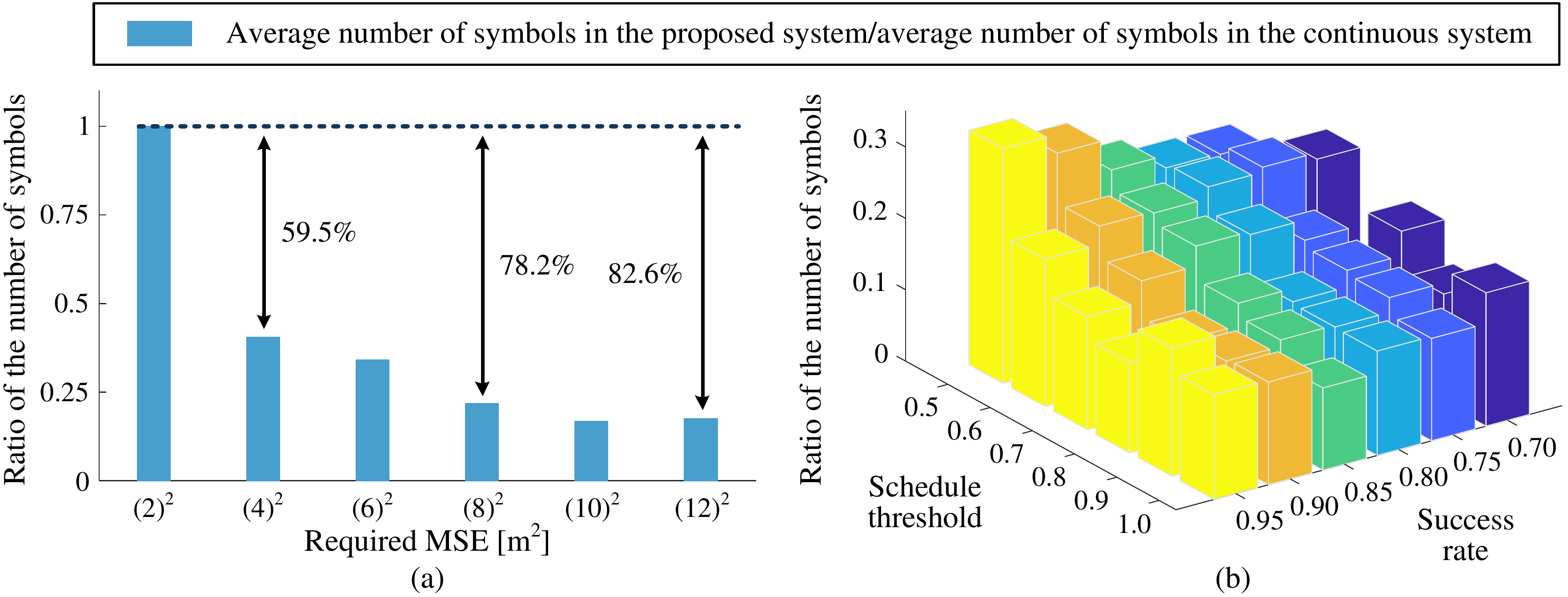}
\caption{Variations of average number of symbols with (a) Required RMSE, (b) success rate and schedule threshold.}
\label{fig_14}
\end{figure}

Then, another experiment is conducted to analyze the influence of parameters $\lambda $ and $P_{Req}^{\left( c \right)}$ on system performance. During the simulation, the value of $P_{Req}^{\left( c \right)}$ increases from 0.70 to 0.95, while $\lambda $ varies from 0.5 to 1.0. The values of other parameters are the same as those described at the beginning of this section. Fig. 12(b), Fig. 13(b) and Fig. 14(b) show the variations of the proposed system's performance metrics with the values of $\lambda $ and $P_{Req}^{\left( c \right)}$. According to Fig. 12(b), the service failure rate of the proposed system decreases rapidly with the increase of $P_{Req}^{\left( c \right)}$ and the decrease of $\lambda $. When $\lambda  \le 0.8$ and $P_{Req}^{\left( c \right)} \ge 0.80$, the service failure rate is lower than $1.435 \times {10^{ - 4}}$, which is negligible for some applications without requirements for high reliability. As can be seen from Fig. 13(b), the average number of scheduling decreases as $\lambda $ and $P_{Req}^{\left( c \right)}$ increase. The average number of scheduling obtained at $\left[ {\lambda ,P_{Req}^{\left( c \right)}} \right] = \left[ {1.0,0.95} \right]$ is 0.1307 times/s, which is only $49.6\% $ of the 0.2633 times/s obtained at $\left[ {0.5,0.7} \right]$. However, it is noteworthy that large values of $\lambda $ may lead to high service failure rates. Thus, in practice, the value of $\lambda $ should be selected carefully according to system's requirements for service reliability and resource efficiency. As shown in Fig. 14(b), the average number of symbols in the proposed system fluctuates with the values of $\lambda $ and $P_{Req}^{\left( c \right)}$, and its variation is quite complicated. The reason for this phenomenon can be intuitively interpreted as follows: Although reducing $P_{Req}^{\left( c \right)}$ could reduce the number of symbols used in each sensing occasion, it also increases the sensing failure rate (Fig. 12(b)), resulting in more additional state sensing (Fig. 13(b)). Therefore, there is no monotonic relationship between the average number of symbols and the value of $P_{Req}^{\left( c \right)}$. The influence of $\lambda $ can be interpreted in the same way. The non-monotonic relationships shown in Fig. 14(b) indicate that it is possible to find a set of $\lambda $ and $P_{Req}^{\left( c \right)}$ that provide the best balance of QoS and resource efficiency, which remains an open question in this article and will be studied in our future work.
\section{Conclusion}
In this article, a novel UAV-enabled positioning system is proposed based on the concept of SCC co-design to provide positioning services that meet UEs' requirements and have high resource efficiency. Different from previous research, we considered the cooperation of sensing, communication and control functionalities in the system design, and studied many issues that have often been overlooked. Specifically, we first established the structure and mathematical models of our system. Then, the UAV stability in open-loop and closed-loop operation modes as well as the theoretical MSE of positioning services in the presence of UAV control errors was derived. It was found that the scheduling of UAV state sensing and blocklength for the transmission of sensing data are two main factors affecting the quality and resource consumption of positioning services. Thus, we further studied the problem of joint sensing scheduling and blocklength allocation, and developed a novel scheme to solve this mixed-integer nonlinear optimization problem efficiently. Numerical results demonstrated that compared with two benchmark systems, the proposed system improves the service reliability and resource efficiency by more than 76.2$\%$ or 82.7$\%$, respectively. We hope this article could lead to a new paradigm for the design of UAV-enabled positioning systems and bring inspiration for the application of SCC co-design concept in UAV systems.

\appendices
\section{Model of TWR Technique}
In the proposed system, the TWR technique is selected to measure inter-UAV distances and the distances between UEs and UAV$_{A}$. In this appendix, we take the inter-UAV distance measurement between the $j$-th UAV$_{A}$ and $i$-th UAV$_{B}$ as an example to establish the mathematical model of TWR. As shown in Fig. 14, the TWR begins with a request message sent by the $i$-th UAV$_{B}$ at local time $t_i^s$. After receiving the request message, the $j$-th UAV$_{A}$ sends back a response message after a fixed response delay ${\tau _D}$ \cite{TWR_Mode}, which will be detected by the UAV$_{B}$ at its local time $\hat t_i^r$. The time interval between the transmission of the request message and the reception of the response message measured by UAV$_{B}$'s local clock can be expressed as \cite{UAV_Geo}
\begin{figure}[!t]
\centering
\includegraphics[height=1.55in,width=3.05in]{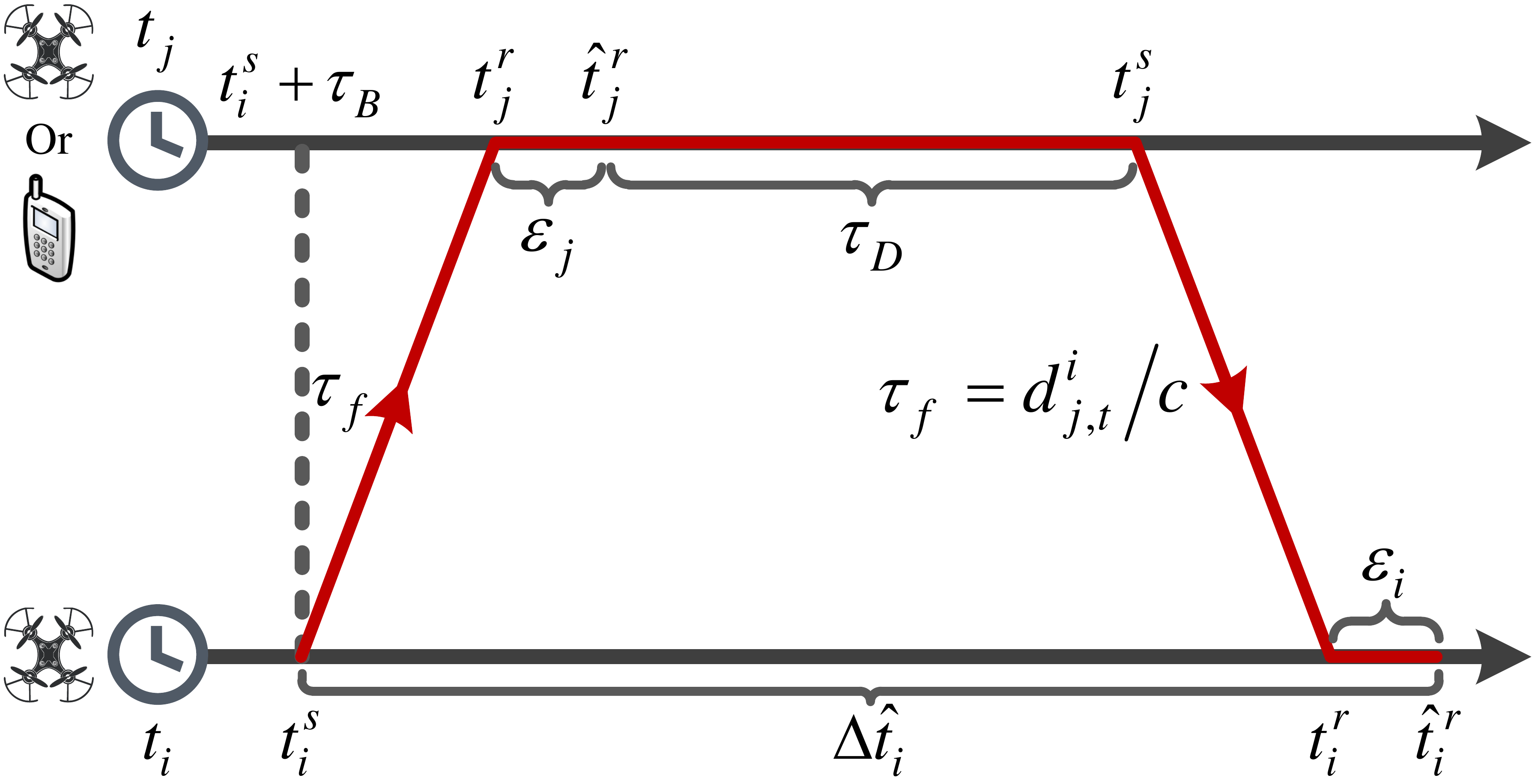}
\caption{Time diagram of TWR.}
\label{fig_15}
\end{figure}
\begin{equation}\tag{53}
\Delta {\hat t_i} = \hat t_i^r - t_i^s = 2{\tau _f}\left( {1 + {\delta _i}} \right) + \left( {{\tau _D} + {\varepsilon _j}} \right) \cdot \frac{{\left( {1 + {\delta _i}} \right)}}{{\left( {1 + {\delta _j}} \right)}} + {\varepsilon _i},
\end{equation}
where ${\tau _f}$ is the time of flight (ToF); ${\delta _j}$ and ${\delta _i}$ denote the clock drifts of the $j$-th UAV$_{A}$ and $i$-th UAV$_{B}$ relative to the reference clock, respectively; ${\varepsilon _j}$ and ${\varepsilon _i}$ are measurement errors caused by UAVs' internal noise.

Then, the estimate of ToF can be written as
\begin{equation}\tag{54}
\begin{split}
{{\hat \tau }_f} = \frac{1}{2}\left( {\Delta {{\hat t}_i} - {\tau _D}} \right) &= {\tau _f}\left( {1 + {\delta _i}} \right) + \frac{{{\tau _D}\left( {{\delta _i} - {\delta _j}} \right) + {\varepsilon _j}\left( {1 + {\delta _i}} \right)}}{{2\left( {1 + {\delta _j}} \right)}} + \frac{{{\varepsilon _i}}}{2}\\
&\approx {\tau _f} + \frac{{{\tau _D}\left( {{\delta _i} - {\delta _j}} \right)}}{2} + \frac{{{\varepsilon _j}}}{2} + \frac{{{\varepsilon _i}}}{2} \approx {\tau _f} + \frac{{{\tau _D}\left( {{\delta _i} - {\delta _j}} \right)}}{2}.
\end{split}
\end{equation}
Please note that in the above equation, the expression of ${\hat \tau _f}$ has been approximated based on the fact that ${\delta _j} \ll 1$ (${\delta _i} \ll 1$) and ${\varepsilon _j} \ll {\tau _D}\left( {{\delta _i} - {\delta _j}} \right)$ (${\varepsilon _i} \ll {\tau _D}\left( {{\delta _i} - {\delta _j}} \right)$) \cite{UAV_Geo}.

Thus, the distance measurement obtained by the TWR technique is given by
\begin{equation}\tag{55}
\hat d_{j,t}^i = c \cdot {\hat \tau _f} = c \cdot {\tau _f} + \frac{{c \cdot {\tau _D}\left( {{\delta _i} - {\delta _j}} \!\right)}}{2} = d_{j,t}^i + n_{d,j,t}^i,
\end{equation}
where $n_{d,j,t}^i = {{c \cdot {\tau _D}\left( {{\delta _i} - {\delta _j}} \right)} \mathord{\left/
 {\vphantom {{c \cdot {\tau _D}\left( {{\delta _i} - {\delta _j}} \right)} 2}} \right.
 \kern-\nulldelimiterspace} 2}$ denotes the distance measurement error caused by clock drifts. Without loss of generality, $n_{d,j,t}^i$ could be modeled as zero-mean Gaussian noise, that is, $n_{d,j,t}^i \sim \left( {0,\sigma _d^2} \right)$ \cite{UAV_Geo,Clk_Mode}.

\section{UAV Dynamics Model}
Since this article is just a preliminary attempt to introduce the SCC co-design concept into UAV-enabled positioning, we simply use the following third-order linear model to describe the dynamics of the $j$-th UAV$_{A}$:
\begin{equation}\tag{56}
\Delta \dot q_{j}^{\left( x \right)} = v_{j}^{\left( x \right)},\qquad\qquad\qquad\quad\;\Delta \dot q_{j}^{\left( y \right)} = v_{j}^{\left( y \right)},\qquad\qquad
\end{equation}
\begin{equation}\tag{57}
\dot v_j^{\left( x \right)} = a_j^{\left( x \right)},\qquad\qquad\qquad\qquad\,\dot v_j^{\left( y \right)} = a_j^{\left( y \right)},\qquad\qquad\;\;\,
\end{equation}
\begin{equation}\tag{58}
\quad\;\,\dot a_j^{\left( x \right)} =  - \frac{1}{\rho }a_j^{\left( x \right)} + \frac{1}{\rho }u_j^{\left( x \right)},\qquad\quad\;\dot a_j^{\left( x \right)} =  - \frac{1}{\rho }a_j^{\left( x \right)} + \frac{1}{\rho }u_j^{\left( x \right)},\quad\,
\end{equation}

The discrete-time version of the above model in the $x$-direction can be written as
\begin{equation}\tag{59}
\Delta q_{j,t + 1}^{\left( x \right)} \!=\! \Delta q_{j,t}^{\left( x \right)} \!+\! \Delta t \cdot v_{j,t}^{\left( x \right)} \!+\! \left( {{\rho ^2}{e^{ - {{\Delta t} \mathord{\left/
 {\vphantom {{\Delta t} \rho }} \right.
 \kern-\nulldelimiterspace} \rho }}} \!+\! \rho \Delta t \!-\! {\rho ^2}} \right)\!a_{j,t}^{\left( x \right)} \!+\! \left[ {{\rho ^2}\!\left(\! { - {e^{ - {{\Delta t} \mathord{\left/
 {\vphantom {{\Delta t} \rho }} \right.
 \kern-\nulldelimiterspace} \rho }}} \!+\! 1} \!\right) + \Delta t\left( { - \rho  \!+\! \Delta t} \right)} \right]u_{j,t}^{\left( x \right)},
\end{equation}
\begin{equation}\tag{60}
v_{j,t + 1}^{\left( x \right)} = v_{j,t}^{\left( x \right)} + \rho \left( {1 - {e^{ - {{\Delta t} \mathord{\left/
 {\vphantom {{\Delta t} \rho }} \right.
 \kern-\nulldelimiterspace} \rho }}}} \right) a_{j,t}^{\left( x \right)} + \left[ {\Delta t + \rho \left( {{e^{ - {{\Delta t} \mathord{\left/
 {\vphantom {{\Delta t} \rho }} \right.
 \kern-\nulldelimiterspace} \rho }}} - 1} \right)} \right] u_{j,t}^{\left( x \right)},
\end{equation}
\begin{equation}\tag{61}
a_{j,t + 1}^{\left( x \right)} = {e^{ - {{\Delta t} \mathord{\left/
 {\vphantom {{\Delta t} \rho }} \right.
 \kern-\nulldelimiterspace} \rho }}}a_{j,t}^{\left( x \right)} + \left( {1 - {e^{ - {{\Delta t} \mathord{\left/
 {\vphantom {{\Delta t} \rho }} \right.
 \kern-\nulldelimiterspace} \rho }}}} \right)u_{j,t}^{\left( x \right)},
\end{equation}
The discrete-time dynamics model in the $y$-direction can be expressed in the same way and will not be repeated here. Then, the expression for matrix ${{\bf{A}}_j}$ in equation (11) is given by
\begin{equation}\tag{62}
{{\bf{A}}_j} = \left[ {\begin{array}{*{20}{c}}
{{{\bf{A}}_{j,1}}}&{{{\bf{A}}_{j,2}}}
\end{array}} \right],
\end{equation}
where
\begin{equation}\tag{63}
{{\bf{A}}_{j,1}} = {\left[ {\begin{array}{*{20}{c}}
1&0&0&0&0&0\\
0&1&0&0&0&0\\
{\Delta t}&0&1&0&0&0\\
0&{\Delta t}&0&1&0&0
\end{array}} \right]^T},
\end{equation}
\begin{equation}\tag{64}
{{\bf{A}}_{j,2}} = \left[ {\begin{array}{*{20}{c}}
{{\rho ^2}{e^{ - {{\Delta t} \mathord{\left/
 {\vphantom {{\Delta t} \rho }} \right.
 \kern-\nulldelimiterspace} \rho }}} + \rho \Delta t - {\rho ^2}}&0\\
0&{{\rho ^2}{e^{ - {{\Delta t} \mathord{\left/
 {\vphantom {{\Delta t} \rho }} \right.
 \kern-\nulldelimiterspace} \rho }}} + \rho \Delta t - {\rho ^2}}\\
{\rho \left( {1 - {e^{ - {{\Delta t} \mathord{\left/
 {\vphantom {{\Delta t} \rho }} \right.
 \kern-\nulldelimiterspace} \rho }}}} \right)}&0\\
0&{\rho \left( {1 - {e^{ - {{\Delta t} \mathord{\left/
 {\vphantom {{\Delta t} \rho }} \right.
 \kern-\nulldelimiterspace} \rho }}}} \right)}\\
{{e^{ - {{\Delta t} \mathord{\left/
 {\vphantom {{\Delta t} \rho }} \right.
 \kern-\nulldelimiterspace} \rho }}}}&0\\
0&{{e^{ - {{\Delta t} \mathord{\left/
 {\vphantom {{\Delta t} \rho }} \right.
 \kern-\nulldelimiterspace} \rho }}}}
\end{array}} \right].
\end{equation}
Similarly, the expression for matrix ${{\bf{B}}_j}$ can be written as
\begin{equation}\tag{65}
{{\bf{B}}_j} = \left[ {\begin{array}{*{20}{c}}
\begin{array}{l}
{\rho ^2}\left( { - {e^{ - {{\Delta t} \mathord{\left/
 {\vphantom {{\Delta t} \rho }} \right.
 \kern-\nulldelimiterspace} \rho }}} + 1} \right)\\
\quad + \Delta t\left( { - \rho  + \Delta t} \right)
\end{array}&0\\
0&\begin{array}{l}
{\rho ^2}\left( { - {e^{ - {{\Delta t} \mathord{\left/
 {\vphantom {{\Delta t} \rho }} \right.
 \kern-\nulldelimiterspace} \rho }}} + 1} \right)\\
\quad + \Delta t\left( { - \rho  + \Delta t} \right)
\end{array}\\
{\Delta t + \rho \left( {{e^{ - {{\Delta t} \mathord{\left/
 {\vphantom {{\Delta t} \rho }} \right.
 \kern-\nulldelimiterspace} \rho }}} - 1} \!\right)}&0\\
0&{\Delta t + \rho \left( {{e^{ - {{\Delta t} \mathord{\left/
 {\vphantom {{\Delta t} \rho }} \right.
 \kern-\nulldelimiterspace} \rho }}} - 1} \right)}\\
{1 - {e^{ - {{\Delta t} \mathord{\left/
 {\vphantom {{\Delta t} \rho }} \right.
 \kern-\nulldelimiterspace} \rho }}}}&0\\
0&{1 - {e^{ - {{\Delta t} \mathord{\left/
 {\vphantom {{\Delta t} \rho }} \right.
 \kern-\nulldelimiterspace} \rho }}}}
\end{array}} \right].
\end{equation}

For the covariance matrix (${{\bf{Q}}_{\bf{w}}}$) of process noise (${{\bf{w}}_{j,t}}$), it can be expressed as \cite{PrN_Cov}
\begin{equation}\tag{66}
{{\bf{Q}}_{\bf{w}}} = E\left\{ {{{\bf{w}}_{j,t}}{\bf{w}}_{j,t}^T} \right\} = \left[ {\begin{array}{*{20}{c}}
{{{\bf{Q}}_{{\bf{w}},1}}}&{{\bf{0}}{}_{4 \times 2}}\\
{{\bf{0}}{}_{2 \times 4}}&{{{\bf{Q}}_{{\bf{w}},2}}}
\end{array}} \right],
\end{equation}
where
\begin{equation}\tag{67}
{{\bf{Q}}_{{\bf{w}},1}} = \left[ {\begin{array}{*{20}{c}}
{{\raise0.7ex\hbox{${\Delta {t^3}\varsigma _x^2}$} \mathord{\left/
 {\vphantom {{\Delta {t^3}\varsigma _x^2} 3}}\right.\kern-\nulldelimiterspace}
\lower0.7ex\hbox{$3$}}}&0&{{\raise0.7ex\hbox{${\Delta {t^2}\varsigma _x^2}$} \mathord{\left/
 {\vphantom {{\Delta {t^2}\varsigma _x^2} 2}}\right.\kern-\nulldelimiterspace}
\lower0.7ex\hbox{$2$}}}&0\\
0&{{\raise0.7ex\hbox{${\Delta {t^3}\varsigma _y^2}$} \mathord{\left/
 {\vphantom {{\Delta {t^3}\varsigma _y^2} 3}}\right.\kern-\nulldelimiterspace}
\lower0.7ex\hbox{$3$}}}&0&{{\raise0.7ex\hbox{${\Delta {t^2}\varsigma _y^2}$} \mathord{\left/
 {\vphantom {{\Delta {t^2}\varsigma _y^2} 2}}\right.\kern-\nulldelimiterspace}
\lower0.7ex\hbox{$2$}}}\\
{{\raise0.7ex\hbox{${\Delta {t^2}\varsigma _x^2}$} \mathord{\left/
 {\vphantom {{\Delta {t^2}\varsigma _x^2} 2}}\right.\kern-\nulldelimiterspace}
\lower0.7ex\hbox{$2$}}}&0&{\Delta t \cdot \varsigma _x^2}&0\\
0&{{\raise0.7ex\hbox{${\Delta {t^2}\varsigma _y^2}$} \mathord{\left/
 {\vphantom {{\Delta {t^2}\varsigma _y^2} 2}}\right.\kern-\nulldelimiterspace}
\lower0.7ex\hbox{$2$}}}&0&{\Delta t \cdot \varsigma _y^2}
\end{array}} \right],
\end{equation}
\begin{equation}\tag{68}
{{\bf{Q}}_{{\bf{w}},2}} = {\rm diag}(\varsigma _x^2,\varsigma _y^2);
\end{equation}
Parameters $\varsigma _x^2$ and $\varsigma _y^2$ represent the continuous-time acceleration process noise intensity in $x$- and $y$-directions, respectively.

\section{Relationship between Blocklength and Transmission Success Rate}
In this article, the A2G link between the $i$-th UAV$_{B}$ and its corresponding BS is modeled as a Rayleigh fading channel, whose channel power gain is represented by ${\left| {{h_{i,t}}} \right|^2}$. Denote the transmit power of UAV$_{B}$ and the noise power at BS as ${P_t}$ and ${N_0}$, then the SNR of the $i$-th UAV$_{B}$'s signal at BS can be expressed as
\begin{equation}\tag{69}
{\gamma _{i,t}} = \left( {{{{P_t}} \mathord{\left/
 {\vphantom {{{P_t}} {{N_0}}}} \right.
 \kern-\nulldelimiterspace} {{N_0}}}} \right) \cdot {\left| {{h_{i,t}}} \right|^2},
\end{equation}

According to \cite{Fin_Len}, if the number of symbols used to transmit the sensing data corresponding to the $j$-th UAV$_{A}$ and $i$-th UAV$_{B}$ in time slot $t$ is $\theta _{j,t}^i$, the success rate of data transmission can be calculated as
\begin{equation}\tag{70}
{P^{\left( s \right)}}\!\left( {\theta _{j,t}^i,{\gamma _{i,t}}} \right) = 1 - Q\left( {f\left( {\theta _{j,t}^i,{\gamma _{i,t}}} \right)} \right),
\end{equation}
where $Q\left( x \right) = {\left( {2\pi } \right)^{{{ - 1} \mathord{\left/
 {\vphantom {{ - 1} 2}} \right.
 \kern-\nulldelimiterspace} 2}}}\int_x^\infty  {{e^{{{ - {z^2}} \mathord{\left/
 {\vphantom {{ - {z^2}} 2}} \right.
 \kern-\nulldelimiterspace} 2}}}dz} $, and
\begin{equation}\tag{71}
f\left( {\theta _{j,t}^i,{\gamma _{i,t}}} \right) = \ln \left( 2 \right)\sqrt {\frac{{\theta _{j,t}^i}}{{{V_{i,t}}}}}  \cdot \left[ {{{\log }_2}\left( {1 + {\gamma _{i,t}}} \right) - \frac{D}{{\theta _{j,t}^i}}} \right];
\end{equation}
$D$ is the size of sensing data (bits) and ${V_{i,t}} = 1 - {\left( {1 + {\gamma _{i,t}}} \right)^{ - 2}}$.

Conversely, for a given success rate $P_{j,t}^{i,\left( s \right)}$ ($P_{j,t}^{i,\left( s \right)} \ge 0.5$), the minimum number of symbols required for data transmission is given by
\begin{equation}\tag{72}
\begin{split}
\theta\! \left(\! {P_{j,t}^{i,\left( s \right)},{\gamma _{i,t}}} \!\right) &= 2D \!\cdot\! \left\{\! {\frac{{{V_{i,t}}}}{D} \!\cdot\! {{\left(\! {\frac{{{Q^{ - 1}}\!\left(\! {1 \!-\! P_{j,t}^{i,\left( s \right)}} \!\right)}}{{\ln \left( 2 \right)}}} \!\right)}^2} \!+\! 2{{\log }_2}\!\left(\! {1 \!+\! {\gamma _{i,t}}} \!\right) \!-\! \left[\! {\frac{{V_{i,t}^2}}{{{D^2}}} \!\cdot\! {{\left(\! {\frac{{{Q^{ - 1}}\!\left(\! {1 \!-\! P_{j,t}^{i,\left( s \right)}} \!\right)}}{{\ln \left( 2 \right)}}} \right)}^4}} \!\right.} \!\right.\\
&{\left. {{{\left. {\left. { + 4{{\log }_2}\!\left(\! 1 \right. \!+\! {\gamma _{i,t}}} \!\right)\!\left(\! {\frac{{{V_{i,t}}}}{D}} \!\right)\!{{\left(\! {\frac{{{Q^{ - 1}}\left(\! {1 \!-\! P_{j,t}^{i,\left( s \right)}} \!\right)}}{{\ln \left( 2 \right)}}} \!\right)}^2}} \right]}^{\frac{1}{2}}}} \!\right\}^{ - 1}},\qquad\qquad\; P_{j,t}^{i,\left( s \right)} \ge 0.5.
\end{split}
\end{equation}

\section{Proof of Proposition 1}
According to equations (43), (44) and (47), $\sigma _{\Delta {\bf{q}},m}^2 \cdot {{\bf{I}}_{{N_A^U}}} - {{\bf{\bar D}}_{\Delta {\bf{q}},m,t + 1}}\left( {{{\bm{\varphi }}_t},{{\bm{\theta }}_t}} \right)$ is a diagonal matrix. Thus, we have
\begin{equation}\tag{73}
\begin{split}
&{\mathop{\rm tr}\nolimits} \left( {{\bf{S}}\left( {{{\bf{p}}_m}} \right)\left( {\sigma _{\Delta {\bf{q}},m}^2 \cdot {{\bf{I}}_{{N_A^U}}} - {{{\bf{\bar D}}}_{\Delta {\bf{q}}.m,t + 1}}\left( {{{\bm{\varphi }}_t},{{\bm{\theta }}_t}} \!\right)} \right){\bf{S}}{{\left( {{{\bf{p}}_m}} \right)}^T}} \right)\\
&\qquad\qquad\qquad\qquad = \sum\limits_{l = 1}^2 {{{\bf{s}}_l}\left( {{{\bf{p}}_m}} \right)\left( {\sigma _{\Delta {\bf{q}},m}^2 \cdot {{\bf{I}}_{{N_A^U}}} - {{{\bf{\bar D}}}_{\Delta {\bf{q}},m,t + 1}}\left( {{{\bm{\varphi }}_t},{{\bm{\theta }}_t}} \right)} \right){{\bf{s}}_l}{{\left( {{{\bf{p}}_m}} \right)}\!^T}} ,
\end{split}
\end{equation}
where ${{\bf{s}}_l}\left( {{{\bf{p}}_m}} \right)$ represents the $l$-th row vector of matrix ${\bf{S}}\left( {{{\bf{p}}_m}} \right)$. Moreover, if condition (46) is satisfied, the elements in diagonal matrix $\sigma _{\Delta {\bf{q}},m}^2 \cdot {{\bf{I}}_{{N_A^U}}} - {{\bf{\bar D}}_{\Delta {\bf{q}},m,t + 1}}\left( {{{\bm{\varphi }}_t},{{\bm{\theta }}_t}} \right)$ are all non-negative, which means that it is a positive semi-definite (PSD) matrix. According to the definition of PSD matrices, we have
\begin{equation}\tag{74}
{{\bf{s}}_l}\left( {{{\bf{p}}_m}} \right)\left( {\sigma _{\Delta {\bf{q}},m}^2 \cdot {{\bf{I}}_{{N_A^U}}} - {{{\bf{\bar D}}}_{\Delta {\bf{q}}.m,t + 1}}\left( {{{\bm{\varphi }}_t},{{\bm{\theta }}_t}} \right)} \right){{\bf{s}}_l}{\left( {{{\bf{p}}_m}} \right)^T} \ge 0.
\end{equation}
Then, the following inequality holds:
\begin{equation}\tag{75}
{\mathop{\rm tr}\nolimits} \left( {{\bf{S}}\left( {{{\bf{p}}_m}} \right)\left( {\sigma _{\Delta {\bf{q}},m}^2 \cdot {{\bf{I}}_{{N_A^U}}} - {{{\bf{\bar D}}}_{\Delta {\bf{q}}.m,t + 1}}\left( {{{\bm{\varphi }}_t},{{\bm{\theta }}_t}} \right)} \right){\bf{S}}{{\left( {{{\bf{p}}_m}} \right)}^T}} \right) \ge 0.
\end{equation}
Finally, we have
\begin{equation}\tag{76}
\begin{split}
{\mathop{\rm tr}\nolimits} &\left( {{\bf{S}}\left( {{{\bf{p}}_m}} \right){{{\bf{\bar D}}}_{\Delta {\bf{q}}.m,t + 1}}\left( {{{\bm{\varphi }}_t},{{\bm{\theta }}_t}} \right){\bf{S}}{{\left( {{{\bf{p}}_m}} \right)}^T}} \right) \le {\mathop{\rm tr}\nolimits} \left( {{\bf{S}}\left( {{{\bf{p}}_m}} \right)\left( {\sigma _{\Delta {\bf{q}},m}^2 \cdot {{\bf{I}}_{{N_A^U}}}} \right){\bf{S}}{{\left( {{{\bf{p}}_m}} \right)}^T}} \right)\\
&\qquad\qquad\qquad\qquad = \sigma _{\Delta {\bf{q}},m}^2 \cdot {\mathop{\rm tr}\nolimits} \left( {{\bf{S}}\left( {{{\bf{p}}_m}} \right){\bf{S}}{{\left( {{{\bf{p}}_m}} \right)}^T}} \right) = \sigma _{\Delta {\bf{q}},m}^2 \cdot {\mathop{\rm tr}\nolimits} \left( {{\bf{P}}\left( {{{\bf{p}}_m}} \right)} \right) = Th{r_m}.
\end{split}
\end{equation}
This completes the proof of Proposition 1.

\bibliographystyle{IEEEtran}
\bibliography{mybib}

\begin{thebibliography}{10}
\providecommand{\url}[1]{#1}
\csname url@samestyle\endcsname
\providecommand{\newblock}{\relax}
\providecommand{\bibinfo}[2]{#2}
\providecommand{\BIBentrySTDinterwordspacing}{\spaceskip=0pt\relax}
\providecommand{\BIBentryALTinterwordstretchfactor}{4}
\providecommand{\BIBentryALTinterwordspacing}{\spaceskip=\fontdimen2\font plus
\BIBentryALTinterwordstretchfactor\fontdimen3\font minus
  \fontdimen4\font\relax}
\providecommand{\BIBforeignlanguage}[2]{{%
\expandafter\ifx\csname l@#1\endcsname\relax
\typeout{** WARNING: IEEEtran.bst: No hyphenation pattern has been}%
\typeout{** loaded for the language `#1'. Using the pattern for}%
\typeout{** the default language instead.}%
\else
\language=\csname l@#1\endcsname
\fi
#2}}
\providecommand{\BIBdecl}{\relax}
\BIBdecl

\bibitem{LBS_Intr}
H.~{Huang}, G.~{Gartner}, J.~M. {Krisp}, M.~{Raubal}, and N.~{Van de Weghe},
  ``Location based services: {Ongoing} evolution and research agenda,''
  \emph{Journal of Location Based Services}, vol.~12, no.~2, pp. 63--93, 2018.

\bibitem{Pos_5G}
R.~Keating, M.~S\"aily, J.~Hulkkonen, and J.~Karjalainen, ``Overview of
  positioning in {5G} new radio,'' in \emph{2019 16th International Symposium
  on Wireless Communication Systems (ISWCS)}, 2019, pp. 320--324.

\bibitem{Pos_6G}
S.~Dang, O.~Amin, B.~Shihada, and M.-S. Alouini, ``What should {6G} be?''
  \emph{Nature Electronics}, vol.~3, no.~1, pp. 20--29, 2020.

\bibitem{GNSS_NLoS}
B.~Xu, Q.~Jia, and L.-T. Hsu, ``Vector tracking loop-based {GNSS NLOS}
  detection and correction: {Algorithm} design and performance analysis,''
  \emph{IEEE Transactions on Instrumentation and Measurement}, vol.~69, no.~7,
  pp. 4604--4619, 2020.

\bibitem{GNSS_Geo}
D.~Lu, S.~Jiang, B.~Cai, W.~Shangguan, K.~Liu, and J.~Luan, ``Quantitative
  analysis of {GNSS} performance under railway obstruction environment,'' in
  \emph{2018 IEEE/ION Position, Location and Navigation Symposium (PLANS)},
  2018, pp. 1074--1080.

\bibitem{Cell_Num}
P.~A. Zandbergen, ``Accuracy of iphone locations: {A} comparison of assisted
  {GPS}, {WiFi} and cellular positioning,'' \emph{Transactions in GIS},
  vol.~13, no.~s1, pp. 5--25, 2009.

\bibitem{Cell_NLoS}
K.~Shamaei and Z.~M. Kassas, ``{LTE} receiver design and multipath analysis for
  navigation in urban environments,'' \emph{NAVIGATION}, vol.~65, no.~4, pp.
  655--675, 2018.

\bibitem{Cell_Geo}
Q.~Liu, R.~Liu, Z.~Wang, and Y.~Zhang, ``Simulation and analysis of device
  positioning in {5G} ultra-dense network,'' in \emph{2019 15th International
  Wireless Communications $\&$ Mobile Computing Conference (IWCMC)}, 2019, pp.
  1529--1533.

\bibitem{UAV_Para}
Y.~Zeng, Q.~Wu, and R.~Zhang, ``Accessing from the sky: {A} tutorial on {UAV}
  communications for {5G} and beyond,'' \emph{Proceedings of the IEEE}, vol.
  107, no.~12, pp. 2327--2375, 2019.

\bibitem{UAV_ABS}
Z.~Wang, L.~Duan, and R.~Zhang, ``Adaptive deployment for {UAV}-aided
  communication networks,'' \emph{IEEE Transactions on Wireless
  Communications}, vol.~18, no.~9, pp. 4531--4543, 2019.

\bibitem{UAV_Relay}
S.~Zeng, H.~Zhang, B.~Di, and L.~Song, ``Trajectory optimization and resource
  allocation for {OFDMA UAV} relay networks,'' \emph{IEEE Transactions on
  Wireless Communications}, pp. 1--1, 2021.

\bibitem{UAV_Data}
M.~Mozaffari, W.~Saad, M.~Bennis, and M.~Debbah, ``Mobile unmanned aerial
  vehicles ({UAVs}) for energy-efficient internet of things communications,''
  \emph{IEEE Transactions on Wireless Communications}, vol.~16, no.~11, pp.
  7574--7589, 2017.

\bibitem{UAV_Pos}
G.~Han, J.~Jiang, C.~Zhang, T.~Q. Duong, M.~Guizani, and G.~K. Karagiannidis,
  ``A survey on mobile anchor node assisted localization in wireless sensor
  networks,'' \emph{IEEE Communications Surveys $\&$ Tutorials}, vol.~18,
  no.~3, pp. 2220--2243, 2016.

\bibitem{UAV_Avai}
Z.~Wang, R.~Liu, Q.~Liu, J.~S. Thompson, and M.~Kadoch, ``Energy-efficient data
  collection and device positioning in {UAV}-assisted {IoT},'' \emph{IEEE
  Internet of Things Journal}, vol.~7, no.~2, pp. 1122--1139, 2020.

\bibitem{UAV_Geo}
Z.~Wang, R.~Liu, Q.~Liu, L.~Han, J.~S. Thompson, Y.~Lin, and W.~Mu, ``Toward
  reliable {UAV}-enabled positioning in mountainous environments: {System}
  design and preliminary results,'' \emph{IEEE Transactions on Reliability},
  pp. 1--29, 2021.

\bibitem{SCC_Sur}
S.~Zhang, R.~P\"ohlmann, E.~Staudinger, and A.~Dammann, ``Assembling a swarm
  navigation system: {Communication}, localization, sensing and control,'' in
  \emph{2021 IEEE 18th Annual Consumer Communications $\&$ Networking
  Conference (CCNC)}, 2021, pp. 1--9.

\bibitem{SCC_Pre}
G.~Zhao, M.~A. Imran, Z.~Pang, Z.~Chen, and L.~Li, ``Toward real-time control
  in future wireless networks: {Communication}-control co-design,'' \emph{IEEE
  Communications Magazine}, vol.~57, no.~2, pp. 138--144, 2019.

\bibitem{SCC_Ben_1}
C.~De~Lima \emph{et~al.}, ``Convergent communication, sensing and localization
  in {6G} systems: {An} overview of technologies, opportunities and
  challenges,'' \emph{IEEE Access}, vol.~9, pp. 26\,902--26\,925, 2021.

\bibitem{SCC_SeCo}
M.~Z. Chowdhury, M.~Shahjalal, S.~Ahmed, and Y.~M. Jang, ``{6G} wireless
  communication systems: {Applications}, requirements, technologies,
  challenges, and research directions,'' \emph{IEEE Open Journal of the
  Communications Society}, vol.~1, pp. 957--975, 2020.

\bibitem{SCC_Ben_2}
Z.~Xiao and Y.~Zeng, ``An overview on integrated localization and communication
  towards {6G},'' \emph{arXiv preprint arXiv:2006.01535}, 2020.

\bibitem{SCC_XUR}
J.~Park, S.~Samarakoon, H.~Shiri, M.~K. Abdel-Aziz, T.~Nishio, A.~Elgabli, and
  M.~Bennis, ``Extreme {URLLC}: {Vision}, challenges, and key enablers,''
  \emph{arXiv preprint arXiv:2001.09683}, 2020.

\bibitem{SCC_UAV}
J.~Zhao, F.~Gao, G.~Ding, T.~Zhang, W.~Jia, and A.~Nallanathan, ``Integrating
  communications and control for {UAV} systems: {Opportunities} and
  challenges,'' \emph{IEEE Access}, vol.~6, pp. 67\,519--67\,527, 2018.

\bibitem{UAV_RSS_1}
H.~Sallouha, M.~M. Azari, A.~Chiumento, and S.~Pollin, ``Aerial anchors
  positioning for reliable {RSS}-based outdoor localization in urban
  environments,'' \emph{IEEE Wireless Communications Letters}, vol.~7, no.~3,
  pp. 376--379, 2018.

\bibitem{UAV_RSS_2}
H.~Sallouha, M.~M. Azari, and S.~Pollin, ``Energy-constrained {UAV} trajectory
  design for ground node localization,'' in \emph{2018 IEEE Global
  Communications Conference (GLOBECOM)}, 2018, pp. 1--7.

\bibitem{UAV_Unc_1}
Z.~Wang, R.~Liu, Q.~Liu, L.~Han, and J.~S. Thompson, ``Feasibility study of
  {UAV}-assisted anti-jamming positioning,'' \emph{IEEE Transactions on
  Vehicular Technology}, pp. 1--1, 2021.

\bibitem{UAV_Unc_2}
Q.~Liu, R.~Liu, Z.~Wang, and J.~S. Thompson, ``{UAV} swarm-enabled localization
  in isolated region: {A} rigidity-constrained deployment perspective,''
  \emph{IEEE Wireless Communications Letters}, pp. 1--1, 2021.

\bibitem{UAV_Unc_3}
Y.~Liu, Y.~Wang, J.~Wang, and Y.~Shen, ``Distributed {3D} relative localization
  of {UAVs},'' \emph{IEEE Transactions on Vehicular Technology}, vol.~69,
  no.~10, pp. 11\,756--11\,770, 2020.

\bibitem{SCC_RAC_1}
F.~Wang and H.~Li, ``Joint power allocation for radar and communication
  co-existence,'' \emph{IEEE Signal Processing Letters}, vol.~26, no.~11, pp.
  1608--1612, 2019.

\bibitem{SCC_RAC_2}
F.~Liu, C.~Masouros, A.~P. Petropulu, H.~Griffiths, and L.~Hanzo, ``Joint radar
  and communication design: {Applications}, state-of-the-art, and the road
  ahead,'' \emph{IEEE Transactions on Communications}, vol.~68, no.~6, pp.
  3834--3862, 2020.

\bibitem{SCC_VCR}
W.~Yuan, F.~Liu, C.~Masouros, J.~Yuan, D.~W.~K. Ng, and N.~Gonz\'alez-Prelcic,
  ``Bayesian predictive beamforming for vehicular networks: {A} low-overhead
  joint radar-communication approach,'' \emph{IEEE Transactions on Wireless
  Communications}, vol.~20, no.~3, pp. 1442--1456, 2021.

\bibitem{SCC_ACC_1}
J.~Mei, K.~Zheng, L.~Zhao, L.~Lei, and X.~Wang, ``Joint radio resource
  allocation and control for vehicle platooning in {LTE-V2V} network,''
  \emph{IEEE Transactions on Vehicular Technology}, vol.~67, no.~12, pp.
  12\,218--12\,230, 2018.

\bibitem{SCC_ACC_2}
A.~Gonz\'alez, N.~Franchi, and G.~Fettweis, ``Control loop aware {LTE-V2X}
  semi-persistent scheduling for string stable {CACC},'' in \emph{2019 IEEE
  30th Annual International Symposium on Personal, Indoor and Mobile Radio
  Communications (PIMRC)}, 2019, pp. 1--7.

\bibitem{SCC_CAC_1}
M.~Eisen, M.~M. Rashid, K.~Gatsis, D.~Cavalcanti, N.~Himayat, and A.~Ribeiro,
  ``Control aware radio resource allocation in low latency wireless control
  systems,'' \emph{IEEE Internet of Things Journal}, vol.~6, no.~5, pp.
  7878--7890, 2019.

\bibitem{SCC_CAC_2}
Y.~Wu, Q.~Yang, H.~Li, K.~S. Kwak, and V.~C.~M. Leung, ``Control-aware
  energy-efficient transmissions for wireless control systems with short
  packets,'' \emph{IEEE Internet of Things Journal}, pp. 1--1, 2021.

\bibitem{Geo_Neg}
K.~C. Ho, X.~Lu, and L.~Kovavisaruch, ``Source localization using {TDOA} and
  {FDOA} measurements in the presence of receiver location errors: {Analysis}
  and solution,'' \emph{IEEE Transactions on Signal Processing}, vol.~55,
  no.~2, pp. 684--696, 2007.

\bibitem{Pos_ILS}
W.~H. FOY, ``Position-location solutions by {Taylor}-series estimation,''
  \emph{IEEE Transactions on Aerospace and Electronic Systems}, vol. AES-12,
  no.~2, pp. 187--194, 1976.

\bibitem{TWR_Mode}
S.~Frattasi and F.~Della~Rosa, \emph{Mobile positioning and tracking: from
  conventional to cooperative techniques}.\hskip 1em plus 0.5em minus
  0.4em\relax John Wiley $\&$ Sons, 2017.

\bibitem{Clk_Mode}
E.~{Akyol}, K.~{Viswanatha}, and K.~{Rose}, ``On conditions for linearity of
  optimal estimation,'' \emph{IEEE Transactions on Information Theory},
  vol.~58, no.~6, pp. 3497--3508, 2012.

\bibitem{PrN_Cov}
S.~Ragothaman, M.~Maaref, and Z.~M. Kassas, ``Multipath-optimal {UAV}
  trajectory planning for urban {UAV} navigation with cellular signals,'' in
  \emph{2019 IEEE 90th Vehicular Technology Conference (VTC2019-Fall)}, 2019,
  pp. 1--6.

\bibitem{Fin_Len}
C.~Pan, H.~Ren, Y.~Deng, M.~Elkashlan, and A.~Nallanathan, ``Joint blocklength
  and location optimization for {URLLC}-enabled {UAV} relay systems,''
  \emph{IEEE Communications Letters}, vol.~23, no.~3, pp. 498--501, 2019.

\end{thebibliography}

\end{document}